\pdfoutput=1
\documentclass[11pt]{article}

\usepackage[margin=1in]{geometry}
\usepackage[T1]{fontenc}
\usepackage[utf8]{inputenc}
\usepackage{lmodern}
\usepackage{microtype}
\usepackage{setspace}
\usepackage{parskip}
\usepackage{amsmath,amssymb,amsfonts,amsthm,mathtools}
\usepackage{aliascnt}
\numberwithin{equation}{section}

\usepackage{graphicx}
\usepackage{subcaption}
\usepackage{booktabs}
\usepackage{array}
\usepackage{longtable}
\usepackage{enumitem}
\usepackage{xcolor}
\usepackage{placeins}
\usepackage{caption}
\usepackage{algorithm}
\usepackage{algpseudocode}
\usepackage{float}
\usepackage{centernot}
\graphicspath{{figures/}}
\setlist[itemize]{leftmargin=1.4em,itemsep=0.15em,topsep=0.15em}
\setlist[enumerate]{leftmargin=1.6em,itemsep=0.15em,topsep=0.15em}

\usepackage[round,authoryear]{natbib}
\usepackage[colorlinks=true,
            linkcolor=blue!50!black,
            citecolor=blue!50!black,
            urlcolor=blue!50!black]{hyperref}
\usepackage[nameinlink,capitalise,noabbrev]{cleveref}

\theoremstyle{plain}
\newtheorem{theorem}{Theorem}[section]
\newaliascnt{proposition}{theorem}
\newtheorem{proposition}[proposition]{Proposition}
\aliascntresetthe{proposition}
\newaliascnt{lemma}{theorem}
\newtheorem{lemma}[lemma]{Lemma}
\aliascntresetthe{lemma}
\newaliascnt{corollary}{theorem}
\newtheorem{corollary}[corollary]{Corollary}
\aliascntresetthe{corollary}

\theoremstyle{definition}
\newaliascnt{definition}{theorem}

\aliascntresetthe{definition}
\newaliascnt{assumption}{theorem}
\newtheorem{assumption}[assumption]{Assumption}
\aliascntresetthe{assumption}
\newaliascnt{example}{theorem}
\newtheorem{example}[example]{Example}
\aliascntresetthe{example}
\theoremstyle{remark}
\newaliascnt{remark}{theorem}
\newtheorem{remark}[remark]{Remark}
\aliascntresetthe{remark}

\crefname{proposition}{Proposition}{Propositions}
\Crefname{proposition}{Proposition}{Propositions}

\crefname{lemma}{Lemma}{Lemmas}
\Crefname{lemma}{Lemma}{Lemmas}

\crefname{corollary}{Corollary}{Corollaries}
\Crefname{corollary}{Corollary}{Corollaries}

\crefname{assumption}{Assumption}{Assumptions}
\Crefname{assumption}{Assumption}{Assumptions}

\crefname{remark}{Remark}{Remarks}
\Crefname{remark}{Remark}{Remarks}

\DeclareMathOperator{\FDP}{FDP}
\DeclareMathOperator{\FDR}{FDR}
\DeclareMathOperator{\Power}{Power}
\DeclareMathOperator{\wBH}{wBH}

\title{\LARGE Certifying a Sufficient Gaussian Dependence Regime for Learned Weighted FDR Control}

\author{Soyeong Oh\\[0.45em] \small Department of Statistics and Probability, Michigan State University \\[0.25em]}

\date{November 2026}

\begin{document}
\maketitle

\begin{abstract}
Learning hypothesis weights from informative auxiliary data can improve the power of multiple testing. Even when the weights are learned independently of the tests, the validity of weighted Benjamini--Hochberg depends on their joint distribution. We use reference observations to certify a sufficient dependence condition for one-sided Gaussian tests. An intersection--union test of uncertain correlation signs determines whether to apply weighted Benjamini--Hochberg or a prespecified fallback at an adjusted level. Under conditional separation of certification and testing, the procedure controls the false discovery rate (FDR) for every Gaussian correlation matrix satisfying the stated sampling assumptions and weight constraints. We establish a boundary obstruction to uniform certification and matching minimax sample-size bounds for worst-case pair sets and the all-pairs problem. In the parameter ranges of the matching bounds, the minimax sample size grows logarithmically with the number of checked pairs and quadratically with the inverse of their positive margin. For homogeneous Gaussian observations with known marginal variances, residual contrasts from the testing subjects provide an independent source of certification scores. Simulations with up to 4,950 checked pairs, comparisons under a common subject budget, and Gaussian designs based on correlations estimated from the Healthy Brain Network illustrate the trade-off between certification and detection power.
\end{abstract}

\section{Introduction}
\label{sec:introduction}

Auxiliary information about which hypotheses are likely to be false can improve the power of multiple testing. Prior weights formalize this idea by allowing rejection thresholds to vary across hypotheses \citep{benjamini1997multiple,genovese2006false}. The weights can also be learned from data \citep{Ignatiadis2016,ignatiadis2021covariate}. Learning weights independently of the testing data addresses one source of selection bias, but it does not determine whether the tests have the dependence structure required by a weighted procedure.

The false discovery rate (FDR) is the expected proportion of false rejections among all rejections, with the proportion defined as zero
when there are none. The Benjamini--Hochberg (BH) step-up procedure controls FDR under independence and under positive regression
dependence on a subset (PRDS)
\citep{benjamini1995controlling,benjamini2001control,sarkar2002some}.
The Benjamini--Yekutieli (BY) correction allows arbitrary dependence at the cost of smaller rejection thresholds \citep{benjamini2001control}.
Their weighted counterparts, wBH and wBY, allow rejection thresholds to vary according to nonnegative hypothesis weights.
The distinction between structured and arbitrary dependence also arises in the $p$-filter of \citet{ramdas2019pfilter}, which
incorporates weighting and dependence-specific reshaping into a framework for multilayer FDR control.

Methods that use auxiliary information include adaptive thresholding and data-driven weighting
\citep{lei2018adapt,li2019multiple,leung2022zap,ignatiadis2021covariate}, as well as independent e-value weights \citep{ignatiadis2024eweights}.
Dependence among tests has been addressed through factor models and estimation of the false discovery proportion (FDP) \citep{sun2009large,friguet2009factor,fan2012estimating,fan2017unknown}.
Conditional calibration uses the joint distribution of the tests to adjust rejection thresholds \citep{fithian2022conditional}.
Symmetrization provides another approach through statistics constructed from independent data splits \citep{du2023symmetrized}.

Recent results establish FDR inflation for BH under two-sided Gaussian tests \citep{dobriban2026gaussianbh} and, more generally, under both one-sided and two-sided Gaussian tests \citep{lei2026gaussiandependence}. Shifted BH procedures modify rejection thresholds using covariance information \citep{sarkar2025shifted,ghosh2025dependence}. In particular, \citet{ghosh2026further} use an independent Wishart covariance estimator to obtain confidence bounds for conditional variances in two-sided Gaussian mean testing. Testing Gaussian multivariate total positivity of order two (MTP$_2$) concerns a different dependence restriction \citep{furmanczyk2023mtp2}. The sample-size scaling in covariance support recovery and large-scale correlation testing is related \citep{jiang2013covariance,cai2016correlations}, although their error criteria differ from those of sign certification.

An alternative to imposing positive dependence conditions is to use e-BH, which controls FDR under arbitrary dependence given valid e-values \citep{wang2022ebh}. Compound e-values provide a broader representation of FDR-controlling procedures \citep{ignatiadis2024compound}, and closure methods can improve e-BH and BY \citep{xu2025closure}. A different statistical problem arises when the weighted testing rule is specified in advance and a sufficient dependence condition for its validity must itself be assessed from reference observations.

We develop a certification procedure for learned weighted BH in one-sided Gaussian testing. Nonnegative correlations involving true nulls suffice for PRDS. Since the true nulls are unknown, we use an intersection--union test of the correlation signs that are not otherwise justified. The certificate selects weighted BH when the sign condition is supported by reference observations and a prespecified fallback otherwise. Conditional separation of certification and testing, together with a bound on erroneous certification, yields finite-sample FDR control for the resulting procedure.

We also characterize the information required for certification. If $d$ correlation signs must be checked and their population correlations are bounded below by a positive margin $\gamma$, a reference sample of order $\gamma^{-2}\log(d/\beta)$ suffices, where $\beta$ is the certification error budget. We establish matching minimax lower bounds for worst-case pair sets and for the all-pairs problem in the stated parameter ranges. A boundary result explains why uniform soundness precludes frequent certification at independence whenever at least one pair is checked. These limits concern certification rather than the validity of the full testing procedure.

Reference scores can be constructed from a separate cohort whose population correlation matrix agrees with that of the testing scores. In a homogeneous Gaussian model with known marginal variances, orthogonal residual contrasts are independent of the sample mean. Consequently, $n$ subjects can supply $n-1$ certification scores without reducing the number of subjects used for testing. We compare this construction with disjoint certification and testing samples under a fixed subject budget. We also establish a power comparison in the completeness regime with Gaussian exponential e-values whose parameters are chosen independently of the testing scores.

FDR control in neuroimaging and inference on brain networks motivate our numerical study \citep{genovese2002thresholding,zalesky2010network}. We examine how often the certificate selects weighted BH, how the required reference sample size changes with dimension, and how the choice of testing branch affects power. We also simulate Gaussian scores using correlations estimated from the Healthy Brain Network (HBN) \citep{alexander2017hbn}. The separate analysis of observed participants documents the construction of these scores and the assumptions that have not been verified.

\section{Setup}
\label{sec:setup}

Let \(X\) denote side information available before certification and testing. Let $\mathcal D_W$, $\mathcal D_C$, and $\mathcal D_T$ denote the data used for family and weight learning, dependence certification, and hypothesis testing, respectively. These data may consist of scores or other transformations of subject-level observations. Using only \((X,\mathcal D_W)\), the procedure may select a finite hypothesis family. If the selected family is empty, take the rejection set to be empty with zero false discovery proportion. All formulas involving division by $m$ below are for $m\ge1$. Otherwise, relabel the selected hypotheses as
\(H_1,\ldots,H_m\), where \(m\ge1\), and define
\(I_0 := \{i\in\{1,\ldots,m\}:H_i\text{ is true}\}\).
Let \(P=(P_1,\ldots,P_m)\in[0,1]^m\) be the corresponding testing $p$-values.
For a rejection set $S\subseteq\{1,\ldots,m\}$, define 
\[\FDP(S)=\frac{|S\cap I_0|}{|S|\vee1}\quad \text{and}\quad \FDR(S)=\mathbb E\{\FDP(S)\}.
\]
When $m>|I_0|$, define $\Power(S)=|S\setminus I_0|/(m-|I_0|)$, the fraction of false nulls rejected.

\subsection{Conditional separation and validity}
\label{sec:three-way-split}

Define $\mathcal H:= \sigma(X,\mathcal D_W).$ Conditional on \(\mathcal H\), the selected and relabeled family, its size \(m\), its null subset \(I_0\), and the learned weights \(W=(W_1,\ldots,W_m)\) are fixed. The certificate \(C\in\{0,1\}\) is measurable with respect to \(\sigma(\mathcal H,\mathcal D_C)\), and
\(P\) is measurable with respect to
\(\sigma(\mathcal H,\mathcal D_T)\).

\begin{assumption}[Conditional routing separation]
\label{ass:routing-separation}
The certificate and testing $p$-values satisfy $C\perp P\mid\mathcal H.$
\end{assumption}

A sufficient implementation condition is \(\mathcal D_C\perp\mathcal D_T\mid\mathcal H\). For example, it follows from
\[
\mathcal L(\mathcal D_W,\mathcal D_C,\mathcal D_T\mid X)
=
\mathcal L(\mathcal D_W\mid X)
\otimes
\mathcal L(\mathcal D_C\mid X)
\otimes
\mathcal L(\mathcal D_T\mid X).
\]
The routing theorem itself requires only
\(C\perp P\mid\mathcal H\).

\begin{assumption}[Conditional null validity]
\label{ass:conditional-null-validity}
For each $i\in I_0$ and $0\le t\le1$, $\Pr(P_i\le t\mid\mathcal H)\le t$ almost surely.
\end{assumption}

\begin{assumption}[Admissible learned weights]
\label{ass:admissible-weights}
The weights are $\mathcal H$-measurable and satisfy $W_i\ge0$ for $i=1,\ldots,m$ and $\sum_{i=1}^m W_i=m$ almost surely.
\end{assumption}

\subsection{Gaussian model}
\label{sec:gaussian-models}

\begin{assumption}[Conditional Gaussian testing model]
\label{ass:gaussian-testing}
The testing score is $Z\mid\mathcal H\sim N_m(\mu,\Gamma)$, where \(\mu\) and \(\Gamma\) are \(\mathcal H\)-measurable and
\(\Gamma\) is a correlation matrix. The hypotheses and $p$-values are
$H_i\colon\mu_i\le0$ and $P_i=1-\Phi(Z_i)$, where $\Phi$ is the standard normal distribution function.
\end{assumption}

Under \cref{ass:gaussian-testing}, for \(i\in I_0\) and \(t\in(0,1)\),
\[
\begin{aligned}
\Pr(P_i\le t\mid\mathcal H)
&=
\Pr\left\{
Z_i\ge\Phi^{-1}(1-t)
\,\middle|\,
\mathcal H
\right\}
\\
&=
1-\Phi\left\{
\Phi^{-1}(1-t)-\mu_i
\right\}
\\
&\le
1-\Phi\left\{
\Phi^{-1}(1-t)
\right\}
\\
&=
t,
\end{aligned}
\]
where the inequality follows from \(\mu_i\le0\).
The cases \(t=0\) and \(t=1\) are immediate. Thus \cref{ass:conditional-null-validity} holds.

\begin{assumption}[Shared-correlation conditional Gaussian certification]
\label{ass:gaussian-certification}
Let $N_C\ge1$ be a positive integer fixed given $\mathcal H$. Conditional on $\mathcal H$, the vectors $Y^{(1)},\ldots,Y^{(N_C)}$ are i.i.d. $N_m(0,\Gamma)$, where \(\Gamma\) is an \(\mathcal H\)-measurable correlation matrix.
When \cref{ass:gaussian-testing} is also imposed, this is the same
\(\Gamma\) that governs \(Z\), and
\((Y^{(1)},\ldots,Y^{(N_C)}) \perp Z \mid\mathcal H\).
\end{assumption}

Define the certification cross-moment matrix by
\begin{equation}\label{eq:certification-matrix}
\widehat\Gamma
=
\frac1{N_C}\sum_{\ell=1}^{N_C}Y^{(\ell)}Y^{(\ell)\top}.
\end{equation}
Under \cref{ass:gaussian-certification}, its conditional expectation is $\Gamma$. The estimator is not renormalized to have unit diagonal.

The shared \(\Gamma\) is an assumption on the population distribution. It does not follow from sample splitting. It must follow from the score construction or an
externally justified sampling model. Here \(N_C\) is the number of independent certification score vectors, not necessarily the number of raw subjects. We use $N_T$ for the number of subjects contributing to the testing score in subject-level constructions and $N_W$ for the number of learning subjects.

\subsection{Idealized Gaussian connectivity construction}
\label{sec:connectivity-instantiation}

Condition on \(\mathcal H\) after a nonempty family has been selected
and relabeled. For one fixed connectivity target, we suppress its label. The resulting argument concerns one selected hypothesis family. Let $V_s
=
(V_{s1},\ldots,V_{sm})^\top$ denote the standardized subject-level connectivity summaries for subject \(s\). For example, \(V_{si}\) may be a standardized Fisher-transformed association between the fixed seed and candidate region \(i\).

Suppose the certification and testing subject-level vectors are
independent conditional on $\mathcal H$, with
$V_s\mid\mathcal H\sim N_m(\theta,\Gamma)$ for every subject,
where $\theta$ and the correlation matrix $\Gamma$ are
$\mathcal H$-measurable. We test $H_i:\theta_i\le0$.
The split and pairing indices and sample sizes are fixed given $\mathcal H$. Locally index the $N_T\ge1$ testing subjects by $s=1,\ldots,N_T$ and define
$Z=N_T^{-1/2}\sum_{s=1}^{N_T}V_s$, so
$Z\mid\mathcal H\sim N_m(\sqrt{N_T}\theta,\Gamma)$.

For certification, use $2N_C$ other subjects, locally relabeled
$V_1,\ldots,V_{2N_C}$, and set
$Y^{(\ell)}=(V_{2\ell}-V_{2\ell-1})/\sqrt2$ for
$\ell=1,\ldots,N_C$. Conditional on $\mathcal H$,
$Y^{(1)},\ldots,Y^{(N_C)}$ are iid $N_m(0,\Gamma)$.

\begin{proposition}[Fixed-target connectivity reduction]
\label{prop:connectivity-reduction}
Under the Gaussian subject-level connectivity model and the three-way subject split above, the testing score satisfies
\cref{ass:gaussian-testing} with \(\mu=\sqrt{N_T}\theta\), while the paired certification scores satisfy \cref{ass:gaussian-certification} with the same correlation matrix \(\Gamma\). Moreover, the paired certification scores and \(Z\) are conditionally
independent given \(\mathcal H\), so every certificate constructed from
the paired scores satisfies \cref{ass:routing-separation}.
\end{proposition}

\begin{proof}
Conditional independence and Gaussian closure give
$\mathbb E(Z\mid\mathcal H)=\sqrt{N_T}\theta$ and
$\operatorname{Cov}(Z\mid\mathcal H)
=N_T^{-1}\sum_{s=1}^{N_T}\Gamma=\Gamma$.
For each certification pair,
$\mathbb E(Y^{(\ell)}\mid\mathcal H)=(\theta-\theta)/\sqrt2=0$
and $\operatorname{Cov}(Y^{(\ell)}\mid\mathcal H)
=\tfrac12(\Gamma+\Gamma)=\Gamma$.
The paired differences are jointly Gaussian and independent across
$\ell$ because they involve disjoint subjects. Their defining subjects
are also disjoint from the testing sample. Hence
$(Y^{(1)},\ldots,Y^{(N_C)})\perp Z\mid\mathcal H$.
Since $C$ is measurable with respect to $\mathcal H$ and the paired
scores, and $P$ with respect to $\mathcal H$ and $Z$, it follows that
$C\perp P\mid\mathcal H$.
\end{proof}

\begin{proposition}[Gaussian certification from the testing subjects]
\label{prop:gaussian-same-subject}
Conditionally on $\mathcal H$, suppose $V_1,\ldots,V_n$ are independent $N_m(\theta,\Gamma)$ observations, where $n\ge2$, $\Gamma$ has unit diagonal, and $\theta,\Gamma$ are $\mathcal H$-measurable. The study observations are separate from the information used to learn the admissible weights in \cref{ass:admissible-weights}. The sample size $n$ and the checked set $J$ are fixed given $\mathcal H$. Let $\bar V=n^{-1}\sum_{s=1}^n V_s$ and $Z=\sqrt n\,\bar V$. Choose a deterministic $(n-1)\times n$ matrix $A$ with $A\mathbf 1_n=0$ and $AA^\top=I_{n-1}$. Here $I_k$ denotes the $k\times k$ identity matrix and $\mathbf1_k$ the vector of $k$ ones. Let $V$ be the $n\times m$ matrix with rows $V_s^\top$, and let $Y^{(\ell)}$ be the transpose of row $\ell$ of $AV$. Then
\[
Z\perp(Y^{(1)},\ldots,Y^{(n-1)})\mid\mathcal H,
\qquad
Y^{(\ell)}\mid\mathcal H\stackrel{\mathrm{iid}}{\sim}N_m(0,\Gamma).
\]
In particular, the sample covariance satisfies
\[
\begin{aligned}
\widehat\Gamma
&=\frac{(AV)^\top(AV)}{n-1}=\frac1{n-1}
\sum_{s=1}^n(V_s-\bar V)(V_s-\bar V)^\top.
\end{aligned}
\]
Moreover,
\[
(n-1)\widehat\Gamma\mid\mathcal H
\sim\operatorname{Wishart}_m(n-1,\Gamma).
\]
Thus the testing score and the certificate based on $\widehat\Gamma$ satisfy \cref{ass:routing-separation} with $N_T=n$ and $N_C=n-1$. The Wishart notation includes singular distributions, defined through sums of Gaussian outer products.
\end{proposition}

\begin{proof}
Condition on $\mathcal H$. The linear transformation that maps the observation matrix $V$ to its normalized row mean $\sqrt n\,\bar V$ and the residual contrasts $AV$ has orthogonal rows in the subject index. The covariance of the transformed observations, stacked by row, is $I_n\otimes\Gamma$. All transformed rows are jointly Gaussian. Their cross-covariances are zero because $A\mathbf1_n=0$. Joint Gaussianity therefore gives independence. The contrast rows have mean zero and covariance $AA^\top\otimes\Gamma=I_{n-1}\otimes\Gamma$. The identity $A^\top A=I_n-\mathbf1_n\mathbf1_n^\top/n$ gives the stated representation of $\widehat\Gamma$. The Wishart distribution and the conditional independence follow.
\end{proof}

With known marginal variances, this construction provides $n-1$ certification scores while testing with all $n$ subjects. Orthogonal contrasts also provide $M_C-1$ scores from a separate cohort of $M_C\ge2$ subjects with a common mean, compared with $\lfloor M_C/2\rfloor$ disjoint paired differences. Pairing is useful if means differ between pairs but agree within each pair. Studentization using the same subjects requires a separate validity argument. Estimated preprocessing need not preserve the assumed population covariance or centering, as discussed in \cref{app:hbn-provenance}.


\section{Certified weighted testing}
\label{sec:method}

Let \(q\in(0,1)\) be the target FDR level and let $0<\beta<q$ be the certificate soundness budget. Define the internal level
\begin{equation}\label{eq:internal-level}
q^\circ=\frac{q-\beta}{1-\beta}.
\end{equation}
Then $q^\circ+(1-q^\circ)\beta=q$.

\subsection{Weighted and fallback branches}
\label{sec:testing-branches}
We use the weighted step-up representation standard in weighted multiple testing \citep{genovese2006false,blanchard2008two}.
For $m\ge1$ and \(w=(w_1,\ldots,w_m)\in[0,\infty)^m\) satisfying
\(\sum_{i=1}^m w_i=m\), \(\alpha\in(0,1)\), and a vector $p\in[0,1]^m$, define
\begin{equation}\label{eq:stepup-count}
K(w,\alpha;p)
:=
\max\left(
\left\{
k\in\{1,\ldots,m\}:
\sum_{i=1}^m
\mathbf 1
\left\{
p_i\le \frac{\alpha w_i k}{m}
\right\}
\ge k
\right\}
\cup\{0\}
\right),
\end{equation}
and
\begin{equation}\label{eq:stepup-rule}
\wBH(w,\alpha;p)
:=
\left\{
i\in\{1,\ldots,m\}\colon
p_i\le
\frac{\alpha w_iK(w,\alpha;p)}{m}
\right\}.
\end{equation}

When the last argument in \eqref{eq:stepup-count} or \eqref{eq:stepup-rule} is omitted, the maps are evaluated at the random testing vector $P$. In particular, $K(w,\alpha)=K(w,\alpha;P)$ and $\wBH(w,\alpha)=\wBH(w,\alpha;P)$. The displayed threshold convention includes a zero $p$-value even at a coordinate with zero weight. Under the Gaussian model, every testing $p$-value is strictly positive almost surely.

\begin{lemma}[Self-consistency]
\label{lem:wbh-self-consistency}
For $m\ge1$, every admissible $w$, and $\alpha\in(0,1)$, $|\wBH(w,\alpha)|=K(w,\alpha).$

Hence 
\[i\in \wBH(w,\alpha)
\quad\Longrightarrow\quad
P_i
\le
\frac{\alpha w_i|\wBH(w,\alpha)|}{m}.
\]

\end{lemma}

\begin{proof}
Let $k=K(w,\alpha),
\;
r=|\wBH(w,\alpha)|.$ By \eqref{eq:stepup-rule},
\[
r
=
\sum_{i=1}^m
\mathbf 1
\left\{
P_i\le
\frac{\alpha w_ik}{m}
\right\}
\ge k.
\]
If \(r>k\), then
\[
\sum_{i=1}^m
\mathbf 1
\left\{
P_i\le
\frac{\alpha w_i r}{m}
\right\}
\ge
\sum_{i=1}^m
\mathbf 1
\left\{
P_i\le
\frac{\alpha w_ik}{m}
\right\}
=r.
\]
Thus \(r\) belongs to the set over which
\(K(w,\alpha)\) is maximized, contradicting
\(r>k\). Therefore \(r=k\), and
\[
i\in \wBH(w,\alpha)
\Longrightarrow
P_i\le
\frac{\alpha w_ik}{m}
=
\frac{\alpha w_i|\wBH(w,\alpha)|}{m}.
\]
\end{proof}

For $m\ge1$, let $c_m=\sum_{r=1}^m1/r$. The weighted branch is $R^w:=\wBH(W,q).$ The two arbitrary-dependence fallbacks are
\begin{equation}\label{eq:fallbacks}
R^{\mathrm{BY}}
:=
\wBH(\mathbf 1_m,q^\circ/c_m),
\qquad
R^{\mathrm{wBY}}
:=
\wBH(W,q^\circ/c_m),
\end{equation}
where $\mathbf1_m$ is the vector of ones.
The first fallback is the harmonic BY correction \citep{benjamini2001control}. The prior-weighted and reshaped formulations are also treated by \citet{ramdas2019pfilter}. The weighted fallback applies the
corresponding harmonic reshaping to the weighted step-up rule \citep{blanchard2008two}.

Before observing \(\mathcal D_C\) or \(\mathcal D_T\), choose the fallback type using at most $\mathcal H$ and set $R^0$ equal to the corresponding rule, either $R^{\mathrm{BY}}$ or $R^{\mathrm{wBY}}$.
These choices define Cert-BY and Cert-wBY, respectively.

\subsection{Gaussian sign certificate}
\label{sec:sign-certificates}

\begin{assumption}[Conditional sign-uncertainty set]
\label{ass:pair-set}
The pair set $J\subseteq\{(i,j):1\le i<j\le m\}$ is $\mathcal H$-measurable. Require $\Gamma_{ij}\ge0$ almost surely for every omitted pair $1\le i<j\le m$ with $(i,j)\notin J$.
\end{assumption}

Thus \(J\) contains every pair whose sign is not justified independently
of \(\mathcal D_C\) and \(\mathcal D_T\). Allowing \(J\) to depend on
\(\mathcal H\) accommodates family selection, but an empirical
correlation computed from \(\mathcal D_W\) does not by itself establish
the omitted-pair sign condition.

Let \(d:=|J|\). For $(i,j)\in J$, the entry $\widehat\Gamma_{ij}$ of the certification cross-moment matrix estimates $\Gamma_{ij}$.
For $d\ge1$, write $s_J(\Gamma)=\min_{(i,j)\in J}\Gamma_{ij}$ and set
\begin{equation}\label{eq:certificate-thresholds}
\begin{aligned}
\tau
&=2\sqrt{\frac{\log(1/\beta)}{N_C}}
  +\frac{2\log(1/\beta)}{N_C},\;
\Delta
=\tau+2\sqrt{\frac{\log(d/\beta)}{N_C}}
  +\frac{2\log(d/\beta)}{N_C}.
\end{aligned}
\end{equation}
For $d\ge1$, the certificate is
\begin{equation}\label{eq:certificate}
C=\mathbf1\left\{\min_{(i,j)\in J}\widehat\Gamma_{ij}\ge\tau\right\}.
\end{equation}
For \(d=0\), set \(C=1\). This is an intersection--union test in the sense of \citet{berger1996bioequivalence}. Setting $C=1$ requires rejection of every component null $\Gamma_{ij}\le0$. False-acceptance control needs no multiplicity correction. The bound for uniform completeness over $J$, by contrast, uses a union bound.

\begin{remark}[Role of the certificate soundness budget]
\label{rem:beta-tradeoff}
The parameter \(\beta\) is prespecified. For fixed \(q\), \(N_C\), and
\(d\ge1\), increasing \(\beta\) decreases \(\tau\), \(\Delta\), and
\(q^\circ\). Certification therefore becomes easier, but the permitted
false-acceptance probability increases and the fallback becomes more
conservative. Decreasing \(\beta\) reverses these effects. We do not tune
\(\beta\) data-adaptively.
\end{remark}

For a desired positive population margin $\gamma>0$, \cref{cor:certification-upper} provides a sufficient choice of $N_C$ that makes $\Delta\le\gamma$. The margin $\gamma$ is used for the sample-size calculation. The decision threshold is $\tau$, whereas $\Delta$ is a sufficient population margin for completeness and is not used to make the certification decision. In particular, the high-acceptance regime may be empty if $\Delta>1$. Nothing in the algorithm assumes that the population margin is known.

The global construction takes $J
=
\{(i,j):1\le i<j\le m\}.$
Externally known exact zeros are one special case of the omitted-pair
condition in \cref{ass:pair-set}. Estimated, small, or nonsignificant
correlations do not justify removing pairs from \(J\). The routed rejection set is
\begin{equation}\label{eq:routed-rule}
R
=
\begin{cases}
R^w,&C=1,\\
R^0,&C=0.
\end{cases}
\end{equation}

\begin{algorithm}[tbp]
\caption{Certified weighted testing}
\label{alg:certified-testing}
\small
\begin{algorithmic}[1]

\Require
target \(q\in(0,1)\), soundness budget \(\beta\in(0,q)\),
\((X,\mathcal D_W,\mathcal D_C,\mathcal D_T)\),
a rule for \(J\) satisfying \cref{ass:pair-set} and a fallback type,
both chosen before certification and testing

\State
Using only \((X,\mathcal D_W)\), select and relabel the hypothesis
family and obtain weights \(W\) satisfying
\(W_i\ge0\) and \(\sum_{i=1}^mW_i=m\).
If \(m=0\), return \(\varnothing\).

\State
Using only \((X,\mathcal D_W)\), construct the pair set \(J\) from the input rule.

\State
Set
$q^\circ$ according to \eqref{eq:internal-level}.

\State
If $J=\varnothing$, set $C\gets1$. Otherwise, compute $\tau$ by \eqref{eq:certificate-thresholds}, the cross-moments by \eqref{eq:certification-matrix}, and $C$ by \eqref{eq:certificate}.

\State
Using \(\mathcal D_T\), compute the testing \(p\)-values and
\(R^w\gets \wBH(W,q)\).

\State
Set $R^0$ to the selected fallback in \eqref{eq:fallbacks}.

\State
\Return $R$ according to \eqref{eq:routed-rule}.

\end{algorithmic}
\end{algorithm}

Given the scores, weights, and checked pairs, the cross-moments require $O(N_Cd)$ arithmetic operations. Sorting the weighted $p$-values for the step-up rules costs $O(m\log m)$, so the total cost of certification and testing is $O(N_Cd+m\log m)$ for $m\ge2$. This excludes the cost of constructing the scores and learning the weights.

\section{Theory}
\label{sec:theory}

Conditioning on \(\mathcal H\) makes the learned weights fixed. The next two branch-validity results follow from standard weighted step-up and dependency-control arguments
\citep{benjamini2001control,genovese2006false,blanchard2008two}.
We state them explicitly because they isolate the assumptions used by the certified procedure.

\subsection{FDR control under routing}
\label{sec:branch-validity}
The next lemma is the harmonic specialization of the reshaping argument used for arbitrary-dependence FDR control by \citet{blanchard2008two}.

\begin{lemma}[Arbitrary-dependence reshaping inequality]
\label{lem:arbitrary-reshaping}
Let $m\ge1$ and let $U\in[0,1]$ be superuniform, so $\Pr(U\le t)\le t$ for $0\le t\le1$.
Let \(K\in\{0,1,\ldots,m\}\) be arbitrary. No independence between
\(U\) and \(K\) is required. Then, for every \(c\ge0\),
\[
\mathbb E
\left[
\frac{\mathbf 1\{U\le c K\}}{K\vee1}
\right]
\le
c c_m.
\]
\end{lemma}

\begin{proof}
Superuniformity at zero gives \(\Pr(U\le0)=0\), so the claim is immediate
when \(c=0\). Suppose \(c>0\). For
\(k=1,\ldots,m\), let
\(A_k := \{c(k-1)<U\le c k\}\).
On \(\{U\le c K\}\cap A_k\), necessarily \(K\ge k\).
Therefore,
\[
\frac{\mathbf 1\{U\le c K\}}{K\vee1}
\le
\sum_{k=1}^m\frac{\mathbf 1(A_k)}{k}
\qquad\text{almost surely}.
\]
Taking expectations and summing by parts,
\[
\begin{aligned}
\sum_{k=1}^m\frac{\Pr(A_k)}{k}
&=
\frac{\Pr(U\le c m)}{m}
+
\sum_{k=1}^{m-1}
\Pr(U\le c k)
\left(
\frac1k-\frac1{k+1}
\right)
\\
&\le
c
+
c
\sum_{k=1}^{m-1}
k
\left(
\frac1k-\frac1{k+1}
\right)
\\
&=
c\sum_{k=1}^m\frac1k
=
c c_m,
\end{aligned}
\]
where
\(\Pr(U\le c k) \le \min(c k,1) \le c k\).
\end{proof}

\begin{proposition}[Validity of the prespecified fallbacks]
\label{prop:fallback-validity}
Suppose $m\ge1$ and \cref{ass:conditional-null-validity,ass:admissible-weights} hold. Then 
\[
\mathbb E
\left\{
\FDP(R^{\mathrm{BY}})
\mid \mathcal H
\right\}
\le q^\circ
\;\text{and}\; 
\mathbb E
\left\{
\FDP(R^{\mathrm{wBY}})
\mid \mathcal H
\right\}
\le q^\circ
\]
almost surely. Consequently, every prespecified choice $R^0\in\{R^{\mathrm{BY}},R^{\mathrm{wBY}}\}$ satisfies
\[
\mathbb E\{\FDP(R^0)\mid\mathcal H\}
\le q^\circ.
\]
\end{proposition}

\begin{proof}
Let $K:=|R^{\mathrm{wBY}}|.$ By \cref{lem:wbh-self-consistency},
\[
i\in R^{\mathrm{wBY}}
\quad\Longrightarrow\quad
P_i
\le
\frac{q^\circ W_iK}{m c_m}.
\]
Hence
\[
\begin{aligned}
\mathbb E
\left\{
\FDP(R^{\mathrm{wBY}})
\mid\mathcal H
\right\}
&=
\sum_{i\in I_0}
\mathbb E
\left[
\frac{\mathbf 1\{i\in R^{\mathrm{wBY}}\}}
{K\vee1}
\,\middle|\,\mathcal H
\right]
\\
&\le
\sum_{i\in I_0}
\mathbb E
\left[
\frac{
\mathbf 1
\left\{
P_i\le
\frac{q^\circ W_iK}{m c_m}
\right\}
}{K\vee1}
\,\middle|\,\mathcal H
\right]
\\
&\le
\frac{q^\circ}{m}
\sum_{i\in I_0}W_i
\\
&\le q^\circ.
\end{aligned}
\]
The second inequality applies \cref{lem:arbitrary-reshaping}
conditionally on \(\mathcal H\), with \(U=P_i\),
\(c=q^\circ W_i/(mc_m)\), and \(K=|R^{\mathrm{wBY}}|\).
Taking \(W_i\equiv1\) gives the corresponding result for
\(R^{\mathrm{BY}}\).
\end{proof}

We use the following cumulative PRDS condition, which is implied by the
PRDS formulation of \citet{benjamini2001control}, also studied by \citet{sarkar2002some}, and is the form
used in the dependency-control argument of \citet{blanchard2008two}.
Conditionally on a sigma-field \(\mathcal H\), the vector
\(P=(P_1,\ldots,P_m)\) satisfies this condition on \(I_0\) if, for every
\(i\in I_0\) and every coordinatewise nondecreasing measurable set
\(D\subseteq[0,1]^m\),
\(t\longmapsto \Pr(P\in D\mid P_i\le t,\mathcal H)\)
is nondecreasing wherever the conditional probability is defined.

\begin{lemma}[PRDS dependency-control inequality]
\label{lem:prds-dependency-control}
Let $m\ge1$, let \(U\in[0,1]\) be superuniform, and let
\(K\in\{0,1,\ldots,m\}\). Suppose that, for each
\(r=1,\ldots,m\),
\(u\longmapsto\Pr(K<r\mid U\le u)\)
is nondecreasing wherever the conditional probability
is defined. Then, for every \(c\ge0\),
\[
\mathbb E\left[
\frac{\mathbf1\{U\le cK\}}{K\vee1}
\right]\le c.
\]
\end{lemma}

\begin{proof}
If \(c=0\), superuniformity gives
\(\Pr(U=0)=0\), so the expectation is zero.
Assume \(c>0\), and set
\(u_r=\min(cr,1),\; r=1,\ldots,m\).
Write \(F(u)=\Pr(U\le u)\).
If \(F(u_m)=0\), then \(F(u_r)=0\) for every
\(r\le m\), and the expectation is again zero.
Otherwise define
\(r_0=\min\{r:F(u_r)>0\}\).

For \(r=r_0,\ldots,m+1\) and \(u\) with \(F(u)>0\),
write
\(A_r(u)=\Pr(K<r\mid U\le u)\).
Then
\(\Pr(K=r\mid U\le u_r) = A_{r+1}(u_r)-A_r(u_r)\).
Since \(U\in[0,1]\),
\(\{U\le cr\}=\{U\le u_r\}\).
The terms with \(r<r_0\) vanish. Therefore
\[
\begin{aligned}
\mathbb E\left[
\frac{\mathbf1\{U\le cK\}}{K\vee1}
\right]
&=
\sum_{r=r_0}^m
\frac1r\Pr(U\le u_r,K=r)
\\
&=
\sum_{r=r_0}^m
\frac{F(u_r)}r
\{A_{r+1}(u_r)-A_r(u_r)\}.
\end{aligned}
\]
Superuniformity gives \(F(u_r)\le u_r\le cr\).
Hence
\[
\mathbb E\left[
\frac{\mathbf1\{U\le cK\}}{K\vee1}
\right]
\le
c\sum_{r=r_0}^m
\{A_{r+1}(u_r)-A_r(u_r)\}.
\]

For the first summand,
\(A_{r_0+1}(u_{r_0})-A_{r_0}(u_{r_0}) \le A_{r_0+1}(u_{r_0})\).
For \(r=r_0+1,\ldots,m\), monotonicity and
\(u_{r-1}\le u_r\) yield
\(A_r(u_{r-1})\le A_r(u_r)\),
so
\(A_{r+1}(u_r)-A_r(u_r) \le A_{r+1}(u_r)-A_r(u_{r-1})\).
Thus
\[
\begin{aligned}
\sum_{r=r_0}^m
\{A_{r+1}(u_r)-A_r(u_r)\}
&\quad\le
A_{r_0+1}(u_{r_0})
+
\sum_{r=r_0+1}^m
\{A_{r+1}(u_r)-A_r(u_{r-1})\}
\\
&\quad=
A_{m+1}(u_m)
=1,
\end{aligned}
\]
because \(K\le m\). Therefore
\[
\mathbb E\left[
\frac{\mathbf1\{U\le cK\}}{K\vee1}
\right]\le c.
\]
\end{proof}

\begin{proposition}[Weighted BH under conditional PRDS]
\label{prop:wbh-prds}
Suppose $m\ge1$ and \cref{ass:admissible-weights} holds. Assume that,
conditionally on \(\mathcal H\),
\[
\Pr(P_i\le t\mid\mathcal H)\le t,
\qquad
i\in I_0,\quad 0\le t\le1,
\]
and \(P\) satisfies the cumulative PRDS condition
stated above on \(I_0\). Then, for every fixed
\(\alpha\in(0,1)\),
\[
\mathbb E\{\FDP(\wBH(W,\alpha))\mid\mathcal H\}
\le\alpha
\]
almost surely.
\end{proposition}

\begin{proof}
Let \(K=|\wBH(W,\alpha)|\). By
\cref{lem:wbh-self-consistency},
\(i\in \wBH(W,\alpha) \Longrightarrow P_i\le\frac{\alpha W_iK}{m}\).

Condition on \(\mathcal H\), so \(W\) is fixed. Write \(K(W,\alpha;p)\) for the step-up index evaluated at \(p\in[0,1]^m\). For \(p\le p'\) coordinatewise and
\(k=1,\ldots,m\),

\[
\sum_{j=1}^m
\mathbf1\left\{
p'_j\le\frac{\alpha W_jk}{m}
\right\}
\le
\sum_{j=1}^m
\mathbf1\left\{
p_j\le\frac{\alpha W_jk}{m}
\right\}.
\]
Every step-up index feasible for \(p'\) is therefore
feasible for \(p\), and
\(K(W,\alpha;p') \le K(W,\alpha;p)\).
By \cref{lem:wbh-self-consistency},
\(|\wBH(W,\alpha;p')| \le|\wBH(W,\alpha;p)|\).
Hence, for each \(r=1,\ldots,m\), the event
\(\{|\wBH(W,\alpha;p)|<r\}\) is coordinatewise
nondecreasing in \(p\).
The cumulative PRDS assumption gives, for every
\(i\in I_0\),
\(u\longmapsto \Pr(K<r\mid P_i\le u,\mathcal H)\)
nondecreasing. Conditional null validity gives
\(\Pr(P_i\le u\mid\mathcal H)\le u\).
Apply \cref{lem:prds-dependency-control}
conditionally on \(\mathcal H\) with
\(U=P_i,\; c=\frac{\alpha W_i}{m}\).
Then
\[
\mathbb E\left[
\frac{\mathbf1\{P_i\le\alpha W_iK/m\}}
{K\vee1}
\,\middle|\,\mathcal H
\right]
\le\frac{\alpha W_i}{m}.
\]
Consequently,
\[
\begin{aligned}
\mathbb E\{\FDP(\wBH(W,\alpha))\mid\mathcal H\}
&=
\sum_{i\in I_0}
\mathbb E\left[
\frac{\mathbf1\{i\in \wBH(W,\alpha)\}}
{K\vee1}
\,\middle|\,\mathcal H
\right]
\\
&\le
\sum_{i\in I_0}
\mathbb E\left[
\frac{\mathbf1\{P_i\le\alpha W_iK/m\}}
{K\vee1}
\,\middle|\,\mathcal H
\right]
\\
&\le
\frac{\alpha}{m}\sum_{i\in I_0}W_i
\le
\frac{\alpha}{m}\sum_{i=1}^mW_i
=\alpha.
\end{aligned}
\]
\end{proof}

\begin{example}[Failure of direct weighted BH under negative correlation]
\label{ex:weighted-bh-inflation}
Let $m=2$, $q=0.1$, and $W=(0.7,1.3)$. Both hypotheses are true. Suppose $Z$ is centered bivariate normal with unit marginal variances and correlation $-0.3$, and set $P_i=1-\Phi(Z_i)$. For $u,v\in[0,1]$, define $F(u,v)=\Pr(P_1\le u,P_2\le v)$. Direct integration yields
\[
\begin{aligned}
\FDR\{\wBH(W,q)\}
&=q+F(0.07,0.13)-F(0.035,0.13)-F(0.07,0.065)\approx0.1005506>q.
\end{aligned}
\]
For $\beta=0.01$, the discounted weighted BY fallback in \eqref{eq:fallbacks} uses $q^\circ/c_2$ with $c_2=3/2$. Its FDR in this example is approximately $0.0608$. A sound certificate accepts with probability at most $\beta$. Conditional separation therefore bounds routed FDR by
\[
0.01(0.1006)+0.99(0.0608)<0.0612<q.
\]
The excess of direct weighted BH over $q$ is small, but it is a strict finite-sample violation. This example illustrates the protective role of routing without asserting that the same violation occurs in the simulations below.
\end{example}

\begin{lemma}[Conditional routing identity]
\label{lem:routing-identity}
Suppose \cref{ass:routing-separation} holds.
Let \(f\) be a bounded real-valued set
functional such that \(f(S)\) is
\(\mathcal H\)-measurable for every fixed
\(S\subseteq\{1,\ldots,m\}\).
Then
\[
\begin{aligned}
\mathbb E\{f(R)\mid\mathcal H\}
&=
\Pr(C=1\mid\mathcal H)
\mathbb E\{f(R^w)\mid\mathcal H\}+
\Pr(C=0\mid\mathcal H)
\mathbb E\{f(R^0)\mid\mathcal H\}
\end{aligned}
\]
almost surely.
\end{lemma}

\begin{proof}
Both \(R^w\) and \(R^0\) are measurable functions of
\((\mathcal H,P)\). Hence \cref{ass:routing-separation} gives $C\perp(R^w,R^0)\mid\mathcal H.$
Moreover,
\(f(R) = f(R^w)\mathbf 1\{C=1\} + f(R^0)\mathbf 1\{C=0\}\).
Taking conditional expectations and applying conditional independence
gives the stated identity.
\end{proof}

Conditional on \(\mathcal H\), the null set \(I_0\) is fixed. Hence
\cref{lem:routing-identity} applies with \(f=\FDP\).

\begin{theorem}[Certified FDR control]
\label{thm:adjusted-certified-fdr}
Let $0<\beta<q<1$ and define $q^\circ$ by \eqref{eq:internal-level}. Suppose \cref{ass:routing-separation} holds and
\[
\mathbb E\{\FDP(R^0)\mid\mathcal H\}\le q^\circ
\]
almost surely. Assume that, for almost every realization of
\(\mathcal H\), either
\[
\mathbb E\{\FDP(R^w)\mid\mathcal H\}\le q
\;\text{or}\;
\Pr(C=1\mid\mathcal H)\le\beta.
\]
Then
\begin{equation}\label{eq:conditional-fdr}
\mathbb E\{\FDP(R)\mid\mathcal H\}\le q
\quad\text{almost surely},
\end{equation}
and consequently $\FDR(R)\le q$.
\end{theorem}

\begin{proof}
Fix a realization of \(\mathcal H\) outside a null set on which the
conditional routing identity and all stated conditional bounds hold.
Write
\(p_C:=\Pr(C=1\mid\mathcal H)\).
If
\(\mathbb E\{\FDP(R^w)\mid\mathcal H\}\le q\),
then \cref{lem:routing-identity} gives
\[
\begin{aligned}
\mathbb E\{\FDP(R)\mid\mathcal H\}
&\le
p_C q+(1-p_C)q^\circ
\\
&\le q.
\end{aligned}
\]
Otherwise, the theorem assumption gives \(p_C\le\beta\). Since
\(0\le\FDP(R^w)\le1\),
\[
\begin{aligned}
\mathbb E\{\FDP(R)\mid\mathcal H\}
&\le
p_C+(1-p_C)q^\circ
\\
&=
q^\circ+(1-q^\circ)p_C
\\
&\le
q^\circ+(1-q^\circ)\beta
=q.
\end{aligned}
\]
Thus
\(\mathbb E\{\FDP(R)\mid\mathcal H\}\le q\; \text{almost surely}\).
Taking expectations completes the proof.
\end{proof}

\begin{theorem}[Boundary obstruction for uniformly sound certification]
\label{thm:boundary-impossibility}
Condition on \(\mathcal H\). Let \(Q_0,Q_1,\ldots\) be conditional laws
of the certification data, with $\|Q-Q'\|_{\mathrm{TV}}=\sup_A|Q(A)-Q'(A)|$ denoting total variation distance. If
\[
Q_r(C=1)\le\beta
\quad\text{for every }r\ge1,
\qquad
\|Q_r-Q_0\|_{\mathrm{TV}}\longrightarrow0,
\]
then
\[
Q_0(C=1)\le\beta.
\]
\end{theorem}

\begin{proof}
For every \(r\ge1\),
\[
Q_0(C=1)
\le
Q_r(C=1)+\|Q_r-Q_0\|_{\mathrm{TV}}
\le
\beta+\|Q_r-Q_0\|_{\mathrm{TV}}.
\]
Letting \(r\to\infty\) proves the claim.
\end{proof}

\subsection{Gaussian certification and sample complexity}
\label{sec:sign-theory}

\begin{proposition}[One-sided Gaussian correlation concentration]
\label{prop:gaussian-correlation-tails}
Under \cref{ass:gaussian-certification,ass:pair-set}, for each checked pair $(i,j)\in J$ and $x>0$,
\[
\Pr\left(
\widehat\Gamma_{ij}-\Gamma_{ij}
\ge
2\sqrt{\frac{x}{N_C}}
+
2\frac{x}{N_C}
\,\middle|\,
\mathcal H
\right)
\le e^{-x},
\]
and
\[
\Pr\left(
\Gamma_{ij}-\widehat\Gamma_{ij}
\ge
2\sqrt{\frac{x}{N_C}}
+
2\frac{x}{N_C}
\,\middle|\,
\mathcal H
\right)
\le e^{-x}.
\]
\end{proposition}

\begin{proof}[Proof of \Cref{prop:gaussian-correlation-tails}]
Condition on \(\mathcal H\) and fix \((i,j)\in J\).
Write \(\rho=\Gamma_{ij}\).
For any $\rho\in[-1,1]$, consider one certification
vector, take independent \(U_1,U_2\sim N(0,1)\)
and write
\[
Y_i=
\frac{
\sqrt{1+\rho}\,U_1+
\sqrt{1-\rho}\,U_2
}{\sqrt2},
\qquad
Y_j=
\frac{
\sqrt{1+\rho}\,U_1-
\sqrt{1-\rho}\,U_2
}{\sqrt2}.
\]
Then
\[
Y_iY_j-\rho
=
\frac{1+\rho}{2}(U_1^2-1)
-
\frac{1-\rho}{2}(U_2^2-1).
\]
For $U\sim N(0,1)$ and $|a|<1/2$,
\[
\begin{aligned}
\log\mathbb E e^{a(U^2-1)}
&=
-a-\frac12\log(1-2a)
\\
&=
\frac12
\sum_{k=2}^{\infty}\frac{(2a)^k}{k}
\\
&\le
\frac12
\sum_{k=2}^{\infty}2^{k-1}|a|^k
\\
&=
\frac{a^2}{1-2|a|}.
\end{aligned}
\]
Hence, for \(|t|<1/2\), independence gives
\[
\begin{aligned}
\log\mathbb E e^{t(Y_iY_j-\rho)}&\le
\frac{(1+\rho)^2t^2/4}
{1-(1+\rho)|t|}
+
\frac{(1-\rho)^2t^2/4}
{1-(1-\rho)|t|}
\\
&\le
\frac{
\{(1+\rho)^2+(1-\rho)^2\}t^2/4
}{1-2|t|}
\\
&=
\frac{(1+\rho^2)t^2/2}{1-2|t|}
\le
\frac{t^2}{1-2|t|}.
\end{aligned}
\]

The bound includes $\rho=\pm1$, since
$1-(1\pm\rho)|t|\ge1-2|t|>0$ for $|t|<1/2$.

For \(0<\lambda<N_C/2\), independence across
certification observations gives
\[
\begin{aligned}
\log\mathbb E
\exp\left\{
\lambda(\widehat\Gamma_{ij}-\Gamma_{ij})
\right\}
&=
\sum_{\ell=1}^{N_C}
\log\mathbb E
\exp\left[
\frac{\lambda}{N_C}
\{Y_i^{(\ell)}Y_j^{(\ell)}-\rho\}
\right]
\\
&\le
N_C
\frac{(\lambda/N_C)^2}
{1-2\lambda/N_C}
\\
&=
\frac{(2/N_C)\lambda^2}
{2\{1-(2/N_C)\lambda\}}.
\end{aligned}
\]

For $x>0$, put $t=\sqrt{x/N_C}$ and
$\lambda=N_Ct/(1+2t)$.
Then \(0<\lambda<N_C/2\). Chernoff's
bound and the preceding MGF inequality give
\[
\begin{aligned}
\log\Pr\left(
\widehat\Gamma_{ij}-\Gamma_{ij}
\ge2t+2t^2
\,\middle|\,\mathcal H
\right)&\le
\frac{\lambda^2}
{N_C(1-2\lambda/N_C)}
-\lambda(2t+2t^2)
\\
&=
\frac{N_Ct^2}{1+2t}
-
\frac{2N_Ct^2(1+t)}{1+2t}
\\
&=-N_Ct^2=-x.
\end{aligned}
\]

The MGF bound is symmetric in the sign of \(t\).
Therefore
\[
\log\mathbb E
\exp\left\{
\lambda(\Gamma_{ij}-\widehat\Gamma_{ij})
\right\}
\le
\frac{(2/N_C)\lambda^2}
{2\{1-(2/N_C)\lambda\}}.
\]
The same Chernoff calculation applied to
$\Gamma_{ij}-\widehat\Gamma_{ij}$ proves the lower-tail bound.
\end{proof}

\begin{corollary}[One-sided Gaussian PRDS]
\label{cor:gaussian-prds}
Under \cref{ass:admissible-weights,ass:gaussian-testing}, suppose $\Gamma_{ij}\ge0$ for every $i\in I_0$ and $j\ne i$. Then, conditionally on \(\mathcal H\), the $p$-value vector \(P\) is
PRDS on \(I_0\). Consequently, $\mathbb E\{\FDP(R^w)\mid\mathcal H\}
\le q.$
\end{corollary}

\begin{proof}
Fix \(i\in I_0\) and let \(D\subseteq[0,1]^m\) be coordinatewise
nondecreasing. For \(t\in(0,1)\),
\(P_i\le t \quad\Longleftrightarrow\quad Z_i\ge\Phi^{-1}(1-t)\).

For \(z\in\mathbb R\), define
\(g_i(z) := \Pr(P\in D\mid Z_i=z,\mathcal H)\).
The conditional law of \(Z_{-i}\) is Gaussian with mean
\(\mu_{-i}+\Gamma_{-i,i}(z-\mu_i)\)
and covariance
\(\Gamma_{-i,-i} - \Gamma_{-i,i}\Gamma_{i,-i}\),
which does not depend on \(z\).
Couple the conditional Gaussian vectors at \(z\le z'\) using the same centered residual.
Then
\(Z_{-i}(z')-Z_{-i}(z) = \Gamma_{-i,i}(z'-z) \ge0\)
coordinatewise. Since \(u\mapsto1-\Phi(u)\) is decreasing, the associated $p$-value vectors satisfy
\(P(z')\le P(z)\).
Because \(D\) is coordinatewise nondecreasing,
\(\mathbf1\{P(z')\in D\} \le \mathbf1\{P(z)\in D\}\).
Therefore \(g_i\) is nonincreasing.
Since \(P_i=1-\Phi(Z_i)\), the map
\(t\longmapsto g_i\{\Phi^{-1}(1-t)\}\)
is a regular conditional version of
\(\Pr(P\in D\mid P_i=t,\mathcal H)\)
for \(0<t<1\).
Both constituent maps are nonincreasing, so their composition is nondecreasing.
This proves standard PRDS on \(I_0\).

For $0<t\le t'\le1$, the conditional distribution of $P_i$ given $P_i\le t$ is stochastically dominated by that given $P_i\le t'$, conditional on $\mathcal H$. Indeed, for $u<t$ its distribution function is $\Pr(P_i\le u\mid\mathcal H)/\Pr(P_i\le t\mid\mathcal H)$, which decreases as $t$ increases; for $u\ge t$ it equals one. Since the conditional probability of $D$ given $P_i=u$ is nondecreasing in $u$, averaging over these truncated laws gives
\[
\Pr(P\in D\mid P_i\le t,\mathcal H)
\le\Pr(P\in D\mid P_i\le t',\mathcal H).
\]
The case $t=0$ is excluded because its conditioning event has probability zero. Thus cumulative PRDS holds. \Cref{prop:wbh-prds} with $\alpha=q$ yields $\mathbb E\{\FDP(R^w)\mid\mathcal H\}\le q$.
\end{proof}

Because \(I_0\) is unknown, the operational certificate verifies a
stronger null-free sign condition on all prespecified certified pairs.
This condition is sufficient, but not necessary, for the validity of the
weighted-BH branch.

\begin{remark}[Sufficiency rather than necessity]
\label{rem:null-free-sign-condition}
The distinction between the null-index condition in
\cref{cor:gaussian-prds} and the operational null-free condition is
substantive. For example, let \(m=3\), \(I_0=\{1\}\), and
\[
\Gamma_a
=
\begin{pmatrix}
1&a&a\\
a&1&-a\\
a&-a&1
\end{pmatrix},
\qquad
0<a<\frac12.
\]
The eigenvalues of \(\Gamma_a\) are
\(1-2a,\; 1+a,\; 1+a\),
so \(\Gamma_a\) is a valid correlation matrix. Moreover,
\((\Gamma_a)_{1j}\ge0\; \text{for every }j\ne1\).
Thus, under the remaining assumptions of
\cref{cor:gaussian-prds}, the $p$-values are PRDS on \(I_0\) and the
weighted-BH branch controls FDR. Nevertheless, for the global pair set,
\(s_J(\Gamma_a)=-a<0\).
Hence a negative certified-pair correlation can arise entirely between
nonnull coordinates. More generally, neither \(s_J(\Gamma)<0\) nor the
sample-level outcome \(C=0\) implies that weighted BH is invalid. It means
only that the chosen sufficient null-free sign class was not certified.
\end{remark}

\begin{theorem}[Intersection--union Gaussian sign certification]
\label{thm:sign-certificate}
Suppose \cref{ass:gaussian-certification,ass:pair-set} hold. Let
\(d=|J|\) and let \(C\) be defined as in
\cref{sec:sign-certificates}, using \eqref{eq:certificate} for $d\ge1$. When \(d\ge1\), let
\(\tau\) and \(\Delta\) be defined by \eqref{eq:certificate-thresholds}, with $s_J(\Gamma)=\min_{(i,j)\in J}\Gamma_{ij}$.

If $d\ge1$, then, almost surely,
\[
\begin{aligned}
s_J(\Gamma)\le0
&\ \Longrightarrow\ \Pr(C=1\mid\mathcal H)\le\beta,\;\
s_J(\Gamma)\ge\Delta
\ \Longrightarrow\ \Pr(C=0\mid\mathcal H)\le\beta.
\end{aligned}
\]
If, in addition,
\cref{ass:admissible-weights,ass:gaussian-testing} hold and \(R^0\) is
a prespecified fallback from \cref{sec:testing-branches}, then, for
every \(d\ge0\),
\[
\mathbb E\{\FDP(R)\mid\mathcal H\}\le q
\quad\text{almost surely},
\]
and consequently $\FDR(R)\le q$.
\end{theorem}

\begin{proof}

Fix a realization of $\mathcal H$ for which the stated assumptions hold. On $\{m=0\}$ the rejection set is empty and the FDR contribution is zero. For $m\ge1$ and $d\ge1$, first use only \cref{ass:gaussian-certification,ass:pair-set}.

If \(s_J(\Gamma)\le0\), choose \((i,j)\in J\) such that
\(\Gamma_{ij}\le0\). Then
\[
\begin{aligned}
\Pr(C=1\mid\mathcal H)
&\le
\Pr(\widehat\Gamma_{ij}\ge\tau\mid\mathcal H)
\\
&\le
\Pr(
\widehat\Gamma_{ij}-\Gamma_{ij}\ge\tau
\mid\mathcal H)
\\
&\le
\beta,
\end{aligned}
\]
where the last inequality follows from
\cref{prop:gaussian-correlation-tails} with
\(x=\log(1/\beta)\).

If \(s_J(\Gamma)\ge\Delta\), then
\[
\begin{aligned}
\{C=0\}
&\subseteq
\bigcup_{(i,j)\in J}
\{\widehat\Gamma_{ij}<\tau\}\subseteq
\bigcup_{(i,j)\in J}
\left\{
\Gamma_{ij}-\widehat\Gamma_{ij}
\ge
\Delta-\tau
\right\}.
\end{aligned}
\]
Since
\(\Delta-\tau = 2\sqrt{\frac{\log(d/\beta)}{N_C}} + 2\frac{\log(d/\beta)}{N_C}\),
\cref{prop:gaussian-correlation-tails} and the union bound give
\(\Pr(C=0\mid\mathcal H) \le d\exp\{-\log(d/\beta)\} = \beta\).

For FDR control, now impose the additional assumptions in the second part of
\cref{thm:sign-certificate}. The certificate is a measurable function of $\mathcal H$ and the reference scores. Since the testing $p$-values are functions of $Z$, the conditional independence in \cref{ass:gaussian-certification} implies \cref{ass:routing-separation}.
For $m\ge1$, \cref{ass:gaussian-testing} implies conditional null validity, so
\cref{prop:fallback-validity} gives \(\mathbb E\{\FDP(R^0)\mid\mathcal H\} \le q^\circ\).

Because $J$ and $\Gamma$ are $\mathcal H$-measurable, the events $\{d=0\}$, $\{d\ge1,\ s_J(\Gamma)\ge0\}$, and $\{d\ge1,\ s_J(\Gamma)<0\}$ form an $\mathcal H$-measurable partition of the nonempty-family event. On the first event all pairs are omitted, so \cref{ass:pair-set} gives nonnegative correlations and $C=1$. On the second event the same sign condition follows from $s_J(\Gamma)\ge0$ and \cref{ass:pair-set}. On both events, \cref{cor:gaussian-prds} yields $\mathbb E\{\FDP(R^w)\mid\mathcal H\}\le q$. On the third event, soundness gives $\Pr(C=1\mid\mathcal H)\le\beta$. Thus the conditional hypotheses of \cref{thm:adjusted-certified-fdr} hold on the entire partition. On the empty-family event, both branches have zero FDP by convention. Applying \cref{thm:adjusted-certified-fdr} yields \eqref{eq:conditional-fdr} and hence $\FDR(R)\le q$.
\end{proof}

\begin{remark}[Exact Gaussian calibration]
\label{rem:exact-gaussian-calibration}
Conditional on $\mathcal H$, for a checked pair with correlation $\rho=\Gamma_{ij}$, the Gaussian cross-moment representation is
\begin{equation}\label{eq:cross-moment-law}
\widehat\Gamma_{ij}\overset{d}{=}
\frac{(1+\rho)U_+-(1-\rho)U_-}{2N_C},
\qquad U_+,U_-\stackrel{\mathrm{iid}}{\sim}\chi^2_{N_C}.
\end{equation}
This follows by summing the independent Gaussian quadratic forms used in the proof of \cref{prop:gaussian-correlation-tails}. Under a common coupling of $U_+$ and $U_-$, the right-hand side is nondecreasing in $\rho$. Hence zero correlation is least favorable for testing $\rho\le0$. For $d\ge1$, let $\tau_{\mathrm{ex}}$ satisfy
\begin{equation}\label{eq:exact-threshold}
\Pr\left(\frac{U_+-U_-}{2N_C}\ge\tau_{\mathrm{ex}}\right)=\beta.
\end{equation}
The cumulative distribution function of $(U_+-U_-)/(2N_C)$ is continuous and strictly increasing on $\mathbb R$, so the quantile in \eqref{eq:exact-threshold} is unique. Replacing $\tau$ by $\tau_{\mathrm{ex}}$ in the intersection--union decision preserves soundness by the same component-test argument. Moreover, $\tau_{\mathrm{ex}}\le\tau$ by \cref{prop:gaussian-correlation-tails}, so the exact-calibrated certificate accepts whenever the original certificate accepts. Under the assumptions of \cref{thm:sign-certificate}, the same FDR guarantee follows. With weighted BY fallback, the routed rejection set can only increase, because the fallback is contained in the weighted BH branch. The explicit threshold in \eqref{eq:certificate-thresholds} is retained throughout the numerical study and the sample-size bounds. The exact threshold provides an alternative calibration under the same Gaussian model.
\end{remark}

An e-value for $H_i$ is a nonnegative statistic with expectation at most one under that null. For $m\ge1$ and nonnegative evidence values $e_1,\ldots,e_m$, e-BH orders them as $e_{(1)}\ge\cdots\ge e_{(m)}$ and rejects the hypotheses with the $k$ largest values, where $k$ is the largest index satisfying $e_{(k)}\ge m/(qk)$. If no such index exists, it rejects none. The procedure controls FDR when the sum of the expectations of the true-null e-values is at most $m$ \citep{wang2022ebh,ignatiadis2024compound}.

\begin{proposition}[Comparison with Gaussian exponential e-values]
\label{prop:exp-evalue-comparison}
Suppose $m\ge1$ and \cref{ass:admissible-weights,ass:gaussian-testing} hold. Let $\lambda_i\ge0$ be $\mathcal H$-measurable and define $E_i=\exp(\lambda_iZ_i-\lambda_i^2/2)$. Let $R_{\mathrm{eBH}}$ be the rejection set from e-BH at level $q$ applied to $W_iE_i$, and let $R^w$ be the rejection set from weighted BH applied to $P_i=1-\Phi(Z_i)$ with the same weights and testing scores. Then
\[
R_{\mathrm{eBH}}\subseteq R^w
\quad\text{for every realization of the scores}.
\]
The weighted exponential e-values are valid for e-BH under arbitrary dependence. Under the further assumptions of \cref{ass:routing-separation,ass:gaussian-certification,ass:pair-set}, with $d\ge1$ and $s_J(\Gamma)\ge\Delta$, let $R$ denote the routed rejection set with weighted BY fallback. If $m>|I_0|$, then
\[
\mathbb E\{\Power(R)\mid\mathcal H\}
\ge(1-\beta)\mathbb E\{\Power(R_{\mathrm{eBH}})\mid\mathcal H\}.
\]
\end{proposition}

\begin{proof}
For each true null, the conditional Gaussian moment generating function gives
$\mathbb E(E_i\mid\mathcal H)=\exp(\lambda_i\mu_i)\le1$.
Hence $\sum_{i\in I_0}\mathbb E(W_iE_i\mid\mathcal H)\le\sum_{i\in I_0}W_i\le m$, which is the weighted e-BH validity condition. Indeed, for a nonempty e-BH rejection set, each rejected null contributes at most $qW_iE_i/m$ to the FDP. Summing over the true nulls and taking conditional expectations gives FDR at most $q$. For a standard normal variable $G$, Markov's inequality applied to $\exp(\lambda_i G)$ gives $1-\Phi(z)\le\exp(-\lambda_i z+\lambda_i^2/2)$ for $\lambda_i\ge0$. Taking $z=Z_i$ yields $P_i\le1/E_i$. If e-BH rejects $k$ hypotheses, each rejected index satisfies $W_iE_i\ge m/(qk)$ and therefore $P_i\le qW_i k/m$. The weighted BH step-up rule can reject at least $k$ hypotheses and includes every e-BH rejection.

Completeness yields $\Pr(C=1\mid\mathcal H)\ge1-\beta$. By \cref{lem:routing-identity} and nonnegativity of power,
\[
\mathbb E\{\Power(R)\mid\mathcal H\}
\ge(1-\beta)\mathbb E\{\Power(R^w)\mid\mathcal H\}
\ge(1-\beta)\mathbb E\{\Power(R_{\mathrm{eBH}})\mid\mathcal H\}.
\]
\end{proof}

The inclusion compares one specified family of exponential e-values. It does not order the power of our procedure against optimized compound e-values, empirical Bayes constructions, or methods based on e-closure \citep{ignatiadis2024compound,xu2025closure}. Those methods may outperform the baseline considered here.

\paragraph{Worst-case certification complexity}

Throughout the Gaussian certification complexity analysis, every supremum over \(\Gamma\) is restricted to symmetric positive-semidefinite matrices with unit diagonal.
Fix $m\ge2$, $\varnothing\ne J\subseteq\{(i,j):1\le i<j\le m\}$, and $0<\gamma\le1$. Let $\Pr_\Gamma$ denote probability under $N_m(0,\Gamma)^{\otimes N_C}$ for $N_C\ge1$. For a measurable certificate \(\phi:(\mathbb R^m)^{N_C}\to\{0,1\}\), write $\phi=1$ for the event that it declares positive signs. We also use $C$ to denote the particular measurable rule defined in \cref{sec:sign-certificates}. Define
\begin{equation}\label{eq:certification-risk}
\begin{aligned}
r_{N_C}(\phi;m,J,\gamma)
:=
\max\Bigg\{&
\sup_{\substack{
\Gamma_{ij}\ge0\text{ for }1\le i<j\le m,\ (i,j)\notin J\\
s_J(\Gamma)<0
}}
\Pr_\Gamma(\phi=1),
\sup_{\substack{
\Gamma_{ij}\ge0\text{ for }1\le i<j\le m,\ (i,j)\notin J\\
s_J(\Gamma)\ge\gamma
}}
\Pr_\Gamma(\phi=0)
\Bigg\}.
\end{aligned}
\end{equation}
For the outer supremum below, \(m\ge2\), \(\binom m2\ge d\), and
\(J\subseteq\{(i,j):1\le i<j\le m\}\)
with \(|J|=d\).
For $d\ge1$ and $0<\beta<1$, define the worst-case minimax certification sample size by
\begin{equation}\label{eq:minimax-size}
\begin{aligned}
N^\star(d,\gamma,\beta)
&=
\inf\Bigl\{N_C\in\mathbb N:\sup_{m,J}
\inf_{\phi\ \mathrm{measurable}}
r_{N_C}(\phi;m,J,\gamma)
\le\beta
\Bigr\}.
\end{aligned}
\end{equation}
Thus \(N^\star\) takes the worst case over admissible ambient dimensions and pair sets of size \(d\). It does not describe the minimax complexity of each fixed \(J\). The adverse class in this risk uses strictly negative signs, whereas the explicit intersection--union certificate controls false acceptance even when the minimum checked correlation is zero.

\begin{corollary}[Sufficient certification sample size]
\label{cor:certification-upper}

In the Gaussian certification benchmark defining \(N^\star\), let \(d\ge1\),
\(0<\gamma\le1\), and \(0<\beta<1\).
If 
\begin{equation}\label{eq:sufficient-size}
N_C\ge
\frac{4\log(d/\beta)}
{\{\sqrt{1+\gamma}-1\}^2},
\end{equation}
then the explicit certificate has risk at
most \(\beta\) uniformly over admissible
ambient dimensions and pair sets of size \(d\).

In particular,
\[
N^\star(d,\gamma,\beta)
\le
\left\lceil
4(1+\sqrt2)^2
\gamma^{-2}\log(d/\beta)
\right\rceil.
\]

\end{corollary}

\begin{proof}
Since $d\ge1$, \eqref{eq:certificate-thresholds} gives
\[
\Delta\le4\sqrt{\frac{\log(d/\beta)}{N_C}}
+4\frac{\log(d/\beta)}{N_C}=4u+4u^2,
\qquad u=\sqrt{\frac{\log(d/\beta)}{N_C}}.
\]
Under \eqref{eq:sufficient-size},
\[
u\le\frac{\sqrt{1+\gamma}-1}{2}
\quad\Longrightarrow\quad \Delta\le4u+4u^2\le\gamma.
\]

Take \(\mathcal H\) to be trivial. The fixed pair set and the outside-\(J\)
sign restriction in the risk class verify \cref{ass:pair-set}.
If \(s_J(\Gamma)<0\), the soundness conclusion of \cref{thm:sign-certificate} gives $\Pr_\Gamma(C=1)
\le
\beta.$ If \(s_J(\Gamma)\ge\gamma\), then
$s_J(\Gamma)
\ge
\gamma
\ge
\Delta,$
and the completeness conclusion of \cref{thm:sign-certificate} gives $\Pr_\Gamma(C=0)
\le
\beta.$ Thus
\(r_{N_C}(C;m,J,\gamma) \le \beta\).
The bound holds for every admissible \(m\) and \(J\) with \(|J|=d\).
Taking the inner infimum over certificates and the outer supremum over
\((m,J)\) in \eqref{eq:minimax-size} proves the stated upper
bound.

Finally, for \(0<\gamma\le1\),
\(\sqrt{1+\gamma}-1 = \frac{\gamma}{\sqrt{1+\gamma}+1} \ge \frac{\gamma}{1+\sqrt2}\).
Therefore
\[
\frac{4\log(d/\beta)}
{\{\sqrt{1+\gamma}-1\}^2}
\le
4(1+\sqrt2)^2
\gamma^{-2}\log(d/\beta).
\]
Since \(N^\star\) is integer-valued,
\(N^\star(d,\gamma,\beta) \le \left\lceil 4(1+\sqrt2)^2 \gamma^{-2}\log(d/\beta) \right\rceil\).

\end{proof}

The lower bound uses a two-point testing argument for the confidence term
and a chi-square mixture argument for the multiplicity term. Standard
versions of these testing reductions are reviewed by
\citet[Chapter~2]{tsybakov2009introduction}.

\begin{theorem}[Certification lower bound, worst case over pair sets]
\label{thm:certification-lower}
There exists a universal constant \(c>0\) such that, for
\(d\ge2,\; 0<\gamma\le\frac16,\; 0<\beta\le\frac1{16}\),
the worst-case certification complexity satisfies
\[
N^\star(d,\gamma,\beta)
\ge
c\gamma^{-2}
\left\{
\log(d+1)+\log(1/\beta)
\right\}.
\]
\end{theorem}

\begin{lemma}[Gaussian divergence identities]
\label{lem:gaussian-divergence-identities}
Let
\[
\Gamma_+
=
\begin{pmatrix}
1&\gamma\\ \gamma&1
\end{pmatrix},
\qquad
\Gamma_-
=
\begin{pmatrix}
1&-\gamma\\ -\gamma&1
\end{pmatrix},
\qquad
0<\gamma<\frac13.
\]
Then
\[
\operatorname{KL}
\{N_2(0,\Gamma_+)\Vert N_2(0,\Gamma_-)\}
=
\frac{2\gamma^2}{1-\gamma^2},
\]
and
\[
1+\chi^2
\{N_2(0,\Gamma_-)\Vert N_2(0,\Gamma_+)\}
=
\sqrt{\frac{1-\gamma^2}{1-9\gamma^2}}.
\]

More generally, let \(\Sigma_0,\Sigma_1\) be \(m\times m\) positive-definite matrices such that \(2\Sigma_1^{-1}-\Sigma_0^{-1}\succ0\).
Then
\[
1+\chi^2
\{N(0,\Sigma_1)\Vert N(0,\Sigma_0)\}
=
|\Sigma_1|^{-1}|\Sigma_0|^{1/2}
\left|
2\Sigma_1^{-1}-\Sigma_0^{-1}
\right|^{-1/2}.
\]
\end{lemma}

\begin{proof}
Direct calculation gives
\(|\Gamma_+|=|\Gamma_-|=1-\gamma^2\),
and
\[
\Gamma_-^{-1}\Gamma_+
=
\frac1{1-\gamma^2}
\begin{pmatrix}
1+\gamma^2&2\gamma\\
2\gamma&1+\gamma^2
\end{pmatrix}.
\]
The Gaussian KL formula therefore yields
\[
\begin{aligned}
\operatorname{KL}
\{N_2(0,\Gamma_+)\Vert N_2(0,\Gamma_-)\}
&=
\frac12\left\{
\operatorname{tr}(\Gamma_-^{-1}\Gamma_+)
-2-\log\frac{|\Gamma_+|}{|\Gamma_-|}
\right\}
\\
&=
\frac{2\gamma^2}{1-\gamma^2}.
\end{aligned}
\]

For centered Gaussian densities \(f_0,f_1\)
with positive definite covariance matrices \(\Sigma_0,\Sigma_1\),
\[
\frac{f_1(x)^2}{f_0(x)}
=
(2\pi)^{-m/2}
|\Sigma_1|^{-1}|\Sigma_0|^{1/2}
\exp\left\{
-\frac12x^\top
(2\Sigma_1^{-1}-\Sigma_0^{-1})x
\right\}.
\]
If $2\Sigma_1^{-1}-\Sigma_0^{-1}$ is positive definite, integrating the Gaussian kernel gives
\[
\begin{aligned}
1+\chi^2(f_1\Vert f_0)
&=
\int\frac{f_1(x)^2}{f_0(x)}\,dx=
|\Sigma_1|^{-1}|\Sigma_0|^{1/2}
\left|
2\Sigma_1^{-1}-\Sigma_0^{-1}
\right|^{-1/2}.
\end{aligned}
\]
In the present construction,
\[
2\Gamma_-^{-1}-\Gamma_+^{-1}
=
\frac1{1-\gamma^2}
\begin{pmatrix}
1&3\gamma\\
3\gamma&1
\end{pmatrix},
\]
and
\(\left| 2\Gamma_-^{-1}-\Gamma_+^{-1} \right| = \frac{1-9\gamma^2}{(1-\gamma^2)^2}\).
The matrix is positive definite for
\(\gamma<1/3\). Consequently,
\[
\begin{aligned}
1+\chi^2
\{N_2(0,\Gamma_-)\Vert N_2(0,\Gamma_+)\}
&=
(1-\gamma^2)^{-1/2}
\left\{
\frac{1-9\gamma^2}{(1-\gamma^2)^2}
\right\}^{-1/2}=
\sqrt{\frac{1-\gamma^2}{1-9\gamma^2}}.
\end{aligned}
\]
\end{proof}

\begin{proof}[Proof of \Cref{thm:certification-lower}]
Take
\(m=2d,\; J=\{(2j-1,2j):j=1,\ldots,d\}\).
Let
\(\Gamma^{(0)} = \operatorname{diag}(\Gamma_+,\ldots,\Gamma_+)\),
and let \(\Gamma^{(j)}\), \(j=1,\ldots,d\),
equal \(\Gamma^{(0)}\) except that its \(j\)th
block is \(\Gamma_-\).

Write
\[
Q_0
=
N_{2d}(0,\Gamma^{(0)})^{\otimes N_C},
\qquad
Q_j
=
N_{2d}(0,\Gamma^{(j)})^{\otimes N_C}.
\]
All these covariance matrices are positive
definite with unit diagonal, and all correlations
outside \(J\) equal zero. Moreover,
\(s_J(\Gamma^{(0)})=\gamma,\; s_J(\Gamma^{(j)})=-\gamma<0, \quad j=1,\ldots,d\).
Thus they belong to the two parameter classes
defining \(r_{N_C}(\phi;2d,J,\gamma)\).

For any measurable certificate \(\phi\)
and \(j\ge1\), tensorization and
\cref{lem:gaussian-divergence-identities} give
\(\operatorname{KL}(Q_0\Vert Q_j) = \frac{2N_C\gamma^2}{1-\gamma^2}\).
The Bretagnolle--Huber inequality yields
\[
\begin{aligned}
Q_0(\phi=0)+Q_j(\phi=1)&\ge
1-\|Q_0-Q_j\|_{\mathrm{TV}}\ge
\frac12
\exp\left\{
-\frac{2N_C\gamma^2}{1-\gamma^2}
\right\}.
\end{aligned}
\]
By \eqref{eq:certification-risk},
\[
\inf_{\phi\ \mathrm{measurable}}
r_{N_C}(\phi;2d,J,\gamma)
\ge
\frac14
\exp\left\{
-\frac{2N_C\gamma^2}{1-\gamma^2}
\right\}.
\]
For \(\gamma\le1/6\) and \(\beta\le1/16\),
\(1-\gamma^2\ge\frac{35}{36},\; \log\frac1{4\beta} \ge\frac12\log\frac1\beta\).
Therefore
\(N_C< \frac{35}{144}\gamma^{-2} \log\frac1\beta\)
implies
\[
\inf_{\phi\ \mathrm{measurable}}
r_{N_C}(\phi;2d,J,\gamma)>\beta.
\]

For the multiplicity term, define
\[
\overline{Q}
=
\frac1d\sum_{j=1}^dQ_j,
\qquad
L_j
=
\frac{\mathrm dQ_j}
{\mathrm dQ_0}.
\]
Under \(Q_0\), the likelihood ratios \(L_j\) depend on disjoint Gaussian blocks. Write $\mathbb E_0$ for expectation under $Q_0$. Hence \(\mathbb E_0(L_j)=1\) and
\(\mathbb E_0(L_jL_k)=1\) for \(j\ne k\).
The divergence identity and tensorization give
\(\mathbb E_0(L_j^2) = \left( \frac{1-\gamma^2}{1-9\gamma^2} \right)^{N_C/2}\).
Consequently,
\[
\begin{aligned}
1+
\chi^2(\overline{Q}\Vert Q_0)&=
\mathbb E_0
\left[
\left(\frac1d\sum_{j=1}^dL_j\right)^2
\right]
\\
&=
\frac1{d^2}
\sum_{j=1}^d\mathbb E_0(L_j^2)
+
\frac1{d^2}
\sum_{j\ne k}\mathbb E_0(L_jL_k)
\\
&=
1+\frac1d
\left[
\left(
\frac{1-\gamma^2}{1-9\gamma^2}
\right)^{N_C/2}-1
\right].
\end{aligned}
\]
Thus
\[
\chi^2(\overline{Q}\Vert Q_0)
=
\frac1d
\left[
\left(
\frac{1-\gamma^2}{1-9\gamma^2}
\right)^{N_C/2}-1
\right].
\]

For \(\gamma\le1/6\),
\[
\begin{aligned}
\log\frac{1-\gamma^2}{1-9\gamma^2}
&\le
-\log(1-9\gamma^2)
\\
&\le
\frac{9\gamma^2}{1-9\gamma^2}
\le12\gamma^2.
\end{aligned}
\]
If
\(N_C<\frac1{6\gamma^2}\log(d+1)\),
then
\(\left( \frac{1-\gamma^2}{1-9\gamma^2} \right)^{N_C/2} <d+1\),
so
\(\chi^2(\overline{Q}\Vert Q_0)<1\).
For every measurable \(\phi\),
\[
\begin{aligned}
2r_{N_C}(\phi;2d,J,\gamma)
&\ge
Q_0(\phi=0)
+
\overline{Q}(\phi=1)
\\
&\ge
1-\|Q_0-\overline{Q}\|_{\mathrm{TV}}
\\
&\ge
1-\frac12
\sqrt{
\chi^2(\overline{Q}\Vert Q_0)
}
\\
&>\frac12.
\end{aligned}
\]
Therefore
\[
\inf_{\phi\ \mathrm{measurable}}
r_{N_C}(\phi;2d,J,\gamma)
\ge
\frac14>\beta.
\]

Finally,
\[
\begin{aligned}
\max\left\{
\frac{35}{144}\log\frac1\beta,
\frac16\log(d+1)
\right\}&\ge
\frac12\left\{
\frac{35}{144}\log\frac1\beta
+
\frac16\log(d+1)
\right\}
\\
&\ge
\frac1{12}
\left\{
\log(d+1)+\log(1/\beta)
\right\}.
\end{aligned}
\]
Consequently, whenever
\(N_C< \frac1{12}\gamma^{-2} \left\{ \log(d+1)+\log(1/\beta) \right\}\),
one of the two risk bounds implies
\[
\inf_{\phi\ \mathrm{measurable}}
r_{N_C}(\phi;2d,J,\gamma)>\beta.
\]
Because this pair-set construction is included
in the outer supremum defining \(N^\star\),
no such \(N_C\) belongs to its defining set.
Thus
\[
N^\star(d,\gamma,\beta)
\ge
\frac1{12}\gamma^{-2}
\left\{
\log(d+1)+\log(1/\beta)
\right\}.
\]
\end{proof}

The preceding construction uses a disjoint matching with \(m=2d\) to
obtain a worst-case result over pair sets. We next treat the all-pairs
choice directly. The proof again uses binary and mixture testing
reductions \citep[Chapter~2]{tsybakov2009introduction} and introduces a
Gaussian common factor similar to the representation in
\citet{AriasCastro2015}. Their problem concerns sparse positive-correlation
alternatives, whereas our positive class requires every certified
correlation to be at least \(\gamma\) and the adverse class contains a
negative pair.

\begin{theorem}[Certification lower bound for the all-pairs set]\label{thm:certification-lower-all-pairs}
There exists a universal constant \(c>0\) such that the following holds.
Let \(m\ge 2\), and set
\(J=\{(i,j):1\le i<j\le m\},\; d=|J|=\binom{m}{2}\).
For
\(0<\gamma\le\frac16,\; 0<\beta\le\frac14\),
every measurable certificate
\(\phi:(\mathbb R^m)^{N_C}\to\{0,1\}\)
satisfying
\(r_{N_C}(\phi;m,J,\gamma)\le\beta\)
must have
\[
N_C
\ge
c\gamma^{-2}
\left\{
\log(d+1)+\log(1/\beta)
\right\}.
\]
\end{theorem}

\begin{proof}
Let
\(k=\left\lfloor\frac m2\right\rfloor,\; \rho=\frac{2\gamma}{1-\gamma}\),
and let \(e_1,\ldots,e_m\) be the standard
basis of \(\mathbb R^m\). Define $G^{(0)}=I_m$ and, for \(j=1,\ldots,k\),
\[
G^{(j)}
=
I_m-\rho
\left(
e_{2j-1}e_{2j}^\top+
e_{2j}e_{2j-1}^\top
\right).
\]
Set
\[
\Gamma^{(j)}
=
\gamma\mathbf1_m\mathbf1_m^\top
+
(1-\gamma)G^{(j)},
\qquad j=0,\ldots,k.
\]
Since \(0<\gamma\le1/6\),
\(0<\rho\le\frac25<1\).
The only nonidentity block of \(G^{(j)}\)
has eigenvalues \(1-\rho\) and \(1+\rho\).
Thus \(G^{(j)}\succ0\). For \(x\ne0\),
\[
x^\top\Gamma^{(j)}x
=
\gamma(\mathbf1_m^\top x)^2
+
(1-\gamma)x^\top G^{(j)}x>0.
\]
Moreover,
\((\Gamma^{(j)})_{rr}=1\; (r=1,\ldots,m)\),
so each \(\Gamma^{(j)}\) is a correlation matrix.

Under \(\Gamma^{(0)}\), every off-diagonal
correlation equals \(\gamma\). Thus
\(s_J(\Gamma^{(0)})=\gamma\).
For \(j\ge1\), the exceptional pair satisfies
\((\Gamma^{(j)})_{2j-1,2j} = \gamma-(1-\gamma)\rho =-\gamma\),
while $(\Gamma^{(j)})_{rs}=\gamma$ for $(r,s)\in J\setminus\{(2j-1,2j)\}$. Symmetry determines the lower triangle. Therefore
\(s_J(\Gamma^{(j)})=-\gamma<0\).

For \(j=0,\ldots,k\), write
\(Q_j = N_m(0,\Gamma^{(j)})^{\otimes N_C}\).
Suppose that a measurable certificate satisfies
\(r_{N_C}(\phi;m,J,\gamma)\le\beta\).
Then
\(Q_0(\phi=0)\le\beta,\; \max_{1\le j\le k}Q_j(\phi=1) \le\beta\).

Let \(\widetilde{Q}_j\) denote the
joint distribution of independent
augmented observations \((U_t,\varepsilon_t)\), $t=1,\ldots,N_C$,
where
\[
U_t\sim N(0,1),
\qquad
\varepsilon_t\sim N_m(0,G^{(j)}),
\qquad
U_t\perp\varepsilon_t.
\]
The common observation map
\[
(U_t,\varepsilon_t)
\longmapsto
\sqrt\gamma\,U_t\mathbf1_m
+
\sqrt{1-\gamma}\,\varepsilon_t
\]
has distribution \(N_m(0,\Gamma^{(j)})\).
Hence \(Q_j\) is the image of
\(\widetilde{Q}_j\) under the same map.

Fix \(j\ge1\). Put
\(a_j=Q_j(\phi=1),\; a_0=Q_0(\phi=1)\).
Then
\(a_j\le\beta,\; a_0\ge1-\beta\).
Data processing through the binary
statistic \(\phi\) gives
\[
\operatorname{KL}(Q_j\Vert Q_0)
\ge
\operatorname{kl}(a_j,a_0),
\]
where
\(\operatorname{kl}(u,v) = u\log\frac{u}{v}+ (1-u)\log\frac{1-u}{1-v}\).
For $0\le u<v<1$, \(\frac{\partial}{\partial v} \operatorname{kl}(u,v) = \frac{v-u}{v(1-v)}>0\).
For \(0<u\le\beta<1/2\),
\[
\frac{\partial}{\partial u}
\operatorname{kl}(u,1-\beta)
=
\log
\frac{u\beta}{(1-\beta)(1-u)}
\le0.
\]
Including the endpoints by continuity,
\[
\operatorname{kl}(a_j,a_0)
\ge
\operatorname{kl}(\beta,1-\beta)
=
(1-2\beta)\log\frac{1-\beta}{\beta}.
\]
Since \(\beta\le1/4\),
\(1-2\beta\ge\frac12,\; 1-\beta\ge\sqrt\beta\),
and consequently
\[
\operatorname{KL}(Q_j\Vert Q_0)
\ge\frac14\log\frac1\beta.
\]

The common observation map and the
data-processing inequality give
\[
\begin{aligned}
\operatorname{KL}(Q_j\Vert Q_0)&\le
\operatorname{KL}
(\widetilde{Q}_j
\Vert\widetilde{Q}_0)
\\
&=
N_C\operatorname{KL}
\{N_m(0,G^{(j)})\Vert N_m(0,I_m)\}.
\end{aligned}
\]
Since \(G^{(j)}\) differs from \(I_m\)
only in its exceptional \(2\times2\)
block,
\(\operatorname{tr}(G^{(j)})=m,\; \det(G^{(j)})=1-\rho^2\).
Therefore
\[
\begin{aligned}
\operatorname{KL}
\{N_m(0,G^{(j)})\Vert N_m(0,I_m)\}&=
\frac12\left\{
\operatorname{tr}(G^{(j)})-m
-\log\det(G^{(j)})
\right\}
\\
&=
-\frac12\log(1-\rho^2).
\end{aligned}
\]
Consequently,
\[
\operatorname{KL}(Q_j\Vert Q_0)
\le
-\frac{N_C}{2}\log(1-\rho^2).
\]

Because
\[
-\log(1-\rho^2)
\le
\frac{\rho^2}{1-\rho^2}
=
\frac{4\gamma^2}{1-2\gamma-3\gamma^2}
\]
and
\(1-2\gamma-3\gamma^2\ge\frac7{12}\),
we have
\(\operatorname{KL}(Q_j\Vert Q_0) \le \frac{24}{7}N_C\gamma^2\).
Combining the two inequalities yields
\[
N_C
\ge
\frac7{96}\gamma^{-2}\log\frac1\beta.
\]

For the multiplicity term, define
\[
\overline{Q}
=
\frac1k\sum_{j=1}^kQ_j,
\qquad
\overline{\widetilde{Q}}
=
\frac1k\sum_{j=1}^k
\widetilde{Q}_j.
\]
The common observation map gives
\[
\chi^2(
\overline{Q}\Vert Q_0
)
\le
\chi^2(
\overline{\widetilde{Q}}
\Vert\widetilde{Q}_0
).
\]
The risk bound implies
\[
Q_0(\phi=1)\ge1-\beta,
\qquad
\overline{Q}(\phi=1)\le\beta.
\]
Consequently,
\[
\begin{aligned}
\frac12
&\le1-2\beta
\\
&\le
\|Q_0-\overline{Q}\|_{\mathrm{TV}}
\\
&\le
\frac12
\sqrt{
\chi^2(
\overline{\widetilde{Q}}
\Vert\widetilde{Q}_0
)
}.
\end{aligned}
\]
Thus
\[
\chi^2(
\overline{\widetilde{Q}}
\Vert\widetilde{Q}_0
)\ge1.
\]

Let
\(L_j = \frac{ \mathrm d\widetilde{Q}_j }{ \mathrm d\widetilde{Q}_0 }\). The common-factor coordinates have the
same distribution under every augmented law, and therefore cancel from the likelihood ratio. Under \(\widetilde{Q}_0\), each \(L_j\) depends only on residual coordinates \(2j-1\) and \(2j\), which are independent across \(j\). In this proof $\mathbb E_0$ denotes expectation under $\widetilde Q_0$. Hence, for \(j\ne\ell\),
\(\mathbb E_0(L_jL_\ell) = \mathbb E_0(L_j)\mathbb E_0(L_\ell) =1\).

The exceptional covariance block is
\[
G_\rho
=
\begin{pmatrix}
1&-\rho\\
-\rho&1
\end{pmatrix}.
\]
Its inverse is
\[
G_\rho^{-1}
=
\frac1{1-\rho^2}
\begin{pmatrix}
1&\rho\\
\rho&1
\end{pmatrix},
\]
so
\[
2G_\rho^{-1}-I_2
=
\frac1{1-\rho^2}
\begin{pmatrix}
1+\rho^2&2\rho\\
2\rho&1+\rho^2
\end{pmatrix}.
\]
The two eigenvalues are
\(\frac{1-\rho}{1+\rho},\; \frac{1+\rho}{1-\rho}\).
Thus \(2G_\rho^{-1}-I_2\succ0\).
Moreover,
\(\det(2G_\rho^{-1}-I_2)=1\).
The remaining coordinates contribute
identity blocks. Consequently,
\(|G^{(j)}|=1-\rho^2,\; \left| 2(G^{(j)})^{-1}-I_m \right|=1\).
By \cref{lem:gaussian-divergence-identities},
\[
1+\chi^2
\{N_m(0,G^{(j)})\Vert N_m(0,I_m)\}
=
\frac1{1-\rho^2}.
\]
Tensorization gives
\(\mathbb E_0(L_j^2) = (1-\rho^2)^{-N_C}\).
It follows that
\[
\begin{aligned}
1+
\chi^2(
\overline{\widetilde{Q}}
\Vert\widetilde{Q}_0
)&=
\mathbb E_0
\left[
\left(
\frac1k\sum_{j=1}^kL_j
\right)^2
\right]
\\
&=
\frac1{k^2}
\left\{
k(1-\rho^2)^{-N_C}+k(k-1)
\right\}
\\
&=
1+
\frac1k\left\{
(1-\rho^2)^{-N_C}-1
\right\}.
\end{aligned}
\]
Since the chi-square divergence is at least one,
\((1-\rho^2)^{-N_C}\ge k+1\).
Taking logarithms,
\(N_C\{-\log(1-\rho^2)\} \ge\log(k+1)\).
Using
\(-\log(1-\rho^2)\le\frac{48}{7}\gamma^2\)
gives
\[
N_C\ge
\frac7{48}\gamma^{-2}\log(k+1).
\]

Since
\(d=\binom m2,\; k=\left\lfloor\frac m2\right\rfloor\),
we have \(m\le2(k+1)\) and \(d+1\le m^2\).
As \(k+1\ge2\),
\[
\begin{aligned}
\log(d+1)
&\le2\log m
\\
&\le2\log\{2(k+1)\}
\\
&\le4\log(k+1).
\end{aligned}
\]
Therefore
\[
N_C\ge
\frac7{192}\gamma^{-2}\log(d+1).
\]

Combining the confidence and multiplicity
bounds and using
\(\max\{a,b\}\ge(a+b)/2\), we obtain
\[
\begin{aligned}
N_C
&\ge
\max\left\{
\frac7{96}\gamma^{-2}\log\frac1\beta,
\frac7{192}\gamma^{-2}\log(d+1)
\right\}
\\
&\ge
\frac7{384}\gamma^{-2}
\left\{
\log(d+1)+\log(1/\beta)
\right\}.
\end{aligned}
\]
This proves the result.
\end{proof}

Under the respective parameter ranges of
\cref{thm:certification-lower,thm:certification-lower-all-pairs}, the rate
\[
\gamma^{-2}
\left\{
\log(d+1)+\log(1/\beta)
\right\}
\]
is necessary both for the worst-case benchmark over pair sets and,
separately, for the all-pairs problem.  The latter result
does not establish the same lower bound for every prespecified pair
set, and neither lower bound is a lower bound for FDR control itself.

\begin{corollary}[Independence boundary]
\label{cor:independence-boundary}
Condition on \(\mathcal H\), let \(m\ge2\), fix \(N_C<\infty\), and let $J$ be any nonempty subset of $\{(i,j):1\le i<j\le m\}$.
Let
\(\phi:(\mathbb R^m)^{N_C}\to\{0,1\}\)
be a measurable certificate based on \(N_C\) iid observations from
\(N_m(0,\Gamma)\). 
The supremum below is over correlation matrices satisfying the omitted-pair condition of \cref{ass:pair-set}. Suppose
\[
\sup_{\substack{
\Gamma \\
s_J(\Gamma)<0
}}
\Pr_\Gamma(\phi=1\mid\mathcal H)
\le\beta.
\]
Then
\[
\Pr_{I_m}(\phi=1\mid\mathcal H)
\le\beta.
\]
\end{corollary}

\begin{proof}
Fix \((i,j)\in J\) and choose
\(-1<\rho_r<0\) with \(\rho_r\uparrow0\).
Set $\Gamma_{\rho_r}
=
I_m+\rho_r
(e_ie_j^\top+e_je_i^\top).$ The eigenvalues of \(\Gamma_{\rho_r}\) are
\(1+\rho_r\), \(1-\rho_r\), and \(1\) with
multiplicity \(m-2\).
Hence \(\Gamma_{\rho_r}\) is a correlation matrix satisfying the omitted-pair condition of \cref{ass:pair-set}.
Moreover,
\(s_J(\Gamma_{\rho_r})<0,\; \det(\Gamma_{\rho_r})=1-\rho_r^2,\; \operatorname{tr}(\Gamma_{\rho_r})=m\).
For one observation,
\[
\begin{aligned}
\operatorname{KL}\left(
N_m(0,\Gamma_{\rho_r})
\Vert N_m(0,I_m)
\right)&=
\frac12\left\{
\operatorname{tr}(\Gamma_{\rho_r})
-m-\log\det(\Gamma_{\rho_r})
\right\}=
-\frac12\log(1-\rho_r^2).
\end{aligned}
\]
Tensorization gives
\[
\begin{aligned}
\operatorname{KL}\left(
N_m(0,\Gamma_{\rho_r})^{\otimes N_C}
\Vert
N_m(0,I_m)^{\otimes N_C}
\right)&=
-\frac{N_C}{2}\log(1-\rho_r^2).
\end{aligned}
\]
By Pinsker's inequality,
\[
\begin{aligned}
&\left\|
N_m(0,\Gamma_{\rho_r})^{\otimes N_C}
-
N_m(0,I_m)^{\otimes N_C}
\right\|_{\mathrm{TV}}\le
\frac12
\sqrt{-N_C\log(1-\rho_r^2)}
\longrightarrow0.
\end{aligned}
\]
Let
\(Q_r=N_m(0,\Gamma_{\rho_r})^{\otimes N_C},\; Q_0=N_m(0,I_m)^{\otimes N_C}\).
Since \(s_J(\Gamma_{\rho_r})<0\), uniform soundness implies
\(Q_r(\phi=1)\le\beta\) for every \(r\).
Applying \cref{thm:boundary-impossibility} to \(Q_r\to Q_0\) gives
\[
Q_0(\phi=1)
=
\Pr_{I_m}(\phi=1\mid\mathcal H)
\le\beta.
\]
\end{proof}

The low acceptance probability at independence is intrinsic to certification that is uniformly sound against arbitrarily small negative correlations.

\begin{corollary}[Comparison with the population-sign branch selector]
\label{cor:branch-selector-comparison}
Under \cref{ass:gaussian-certification,ass:pair-set}, let \(d\ge1\),
and let \(R^w\) and \(R^0\) be the two testing branches defined in
\cref{sec:testing-branches}. Define
\begin{equation}\label{eq:population-selector}
R^\star
=
\begin{cases}
R^w,&s_J(\Gamma)\ge0,\\
R^0,&s_J(\Gamma)<0.
\end{cases}
\end{equation}
For every \([0,1]\)-valued set functional \(f\) such that \(f(S)\) is
\(\mathcal H\)-measurable for every fixed
\(S\subseteq\{1,\ldots,m\}\),
\[
\left|
\mathbb E\{f(R)\mid\mathcal H\}
-
\mathbb E\{f(R^\star)\mid\mathcal H\}
\right|
\le
\beta
\]
almost surely on
\(\{s_J(\Gamma)<0\} \cup \{s_J(\Gamma)\ge\Delta\}\).
\end{corollary}

\begin{proof}
Condition on $\mathcal H$ and restrict to the stated comparison region. By \eqref{eq:routed-rule} and \eqref{eq:population-selector}, the two rejection sets coincide whenever $C=\mathbf1\{s_J(\Gamma)\ge0\}$. Since $f$ takes values in $[0,1]$,
\[
\begin{aligned}
\left|\mathbb E\{f(R)-f(R^\star)\mid\mathcal H\}\right|
&\le \mathbb E\{|f(R)-f(R^\star)|\mid\mathcal H\}\\
&\le \Pr(R\ne R^\star\mid\mathcal H)\\
&\le \Pr(C\ne\mathbf1\{s_J(\Gamma)\ge0\}\mid\mathcal H).
\end{aligned}
\]
The last probability is at most $\beta$ by soundness when $s_J(\Gamma)<0$ and by completeness when $s_J(\Gamma)\ge\Delta$, using \cref{thm:sign-certificate}.
\end{proof}

The population-sign branch selector \(R^\star\) is defined for every
value of \(s_J(\Gamma)\), but
\cref{cor:branch-selector-comparison} controls its discrepancy from the
routed rule only when
\(s_J(\Gamma)<0\; \text{or}\; s_J(\Gamma)\ge\Delta\).
We refer to this set as the comparison region. In the transition region
\(0\le s_J(\Gamma)<\Delta\), the selector remains a descriptive
comparator without the corollary's guarantee.

\begin{remark}[Cert-wBY decomposition relative to direct wBY]
\label{rem:cert-wby-gain-cost}
Suppose \cref{ass:routing-separation,ass:admissible-weights} hold and $m\ge1$. Set \(R^0=R^{\mathrm{wBY}}\), and let \(f\)
be as in \cref{lem:routing-identity} and
nondecreasing under set inclusion.
Let
$R^{\mathrm{dir}}=\wBH(W,q/c_m)$
be the direct weighted BY rejection set at level $q$.
The routing identity gives, almost surely,
\[
\begin{aligned}
\mathbb E\{f(R)-f(R^{\mathrm{dir}})\mid\mathcal H\}&=
\Pr(C=1\mid\mathcal H)
\mathbb E\{f(R^w)-f(R^{\mathrm{dir}})\mid\mathcal H\}\\
&\qquad-
\Pr(C=0\mid\mathcal H)
\mathbb E\{f(R^{\mathrm{dir}})-f(R^0)\mid\mathcal H\}.
\end{aligned}
\]
Since \(q^\circ<q\) and \(c_m\ge1\),
\(\frac{q^\circ}{c_m} \le \frac q{c_m} \le q\).
For fixed weights \(W\), monotonicity
in the testing level gives
\[
\alpha\le\alpha'
\quad\Longrightarrow\quad
\wBH(W,\alpha)\subseteq \wBH(W,\alpha').
\]
For each $k\in\{1,\ldots,m\}$,
\[
\sum_{i=1}^m
\mathbf1\left\{
P_i\le\frac{\alpha W_i k}{m}
\right\}
\le
\sum_{i=1}^m
\mathbf1\left\{
P_i\le\frac{\alpha' W_i k}{m}
\right\},
\]
so
\(K(W,\alpha) \le K(W,\alpha')\).
Consequently,
\[
R^0
=
\wBH(W,q^\circ/c_m)
\subseteq
\wBH(W,q/c_m)
=
R^{\mathrm{dir}}
\subseteq
\wBH(W,q)
=
R^w.
\]
Thus both differences are nonnegative. Under \cref{ass:gaussian-certification,ass:pair-set}, if \(d\ge1\) and \(s_J(\Gamma)\ge\Delta\),
\cref{thm:sign-certificate} gives
\(\Pr(C=0\mid\mathcal H)\le\beta\),
which bounds the coefficient of the fallback-cost term. Moreover, when $m>|I_0|$, the same inclusion and routing identity give $0\le\mathbb E\{\Power(R^w)-\Power(R)\mid\mathcal H\}\le\beta$.
No analogous ordering follows for \(\FDP\), which is not monotone under set inclusion.
\end{remark}


\section{Experiments}
\label{sec:experiments}

We examine how the procedure behaves near the sign boundary, how the certificate responds to changes in sample size and dimension, and how much power can be gained when all subjects contribute to testing. We then consider a Gaussian model using correlations estimated from HBN. Throughout, $q=0.10$ and $\beta=0.01$. Except where the comparison explicitly changes subject allocation, competing methods use the same testing $p$-values and weights within each replication. Mean FDP estimates FDR, and mean power estimates the expected fraction of false nulls rejected. We report simultaneous 95\% Hoeffding--Bonferroni (HB) intervals for prespecified families of bounded Monte Carlo means and a Clopper--Pearson (CP) interval for HBN Gaussian certificate acceptance. The interval half-width is given by \eqref{eq:hb-halfwidth}. Full designs, comparator definitions, and computational details appear in \cref{app:additional-experiments}.

\paragraph{Routing transition.}
The first experiment tracks end-to-end routing along a common Gaussian
dependence path with the all-pairs choice of \(J\). Learning,
certification, and testing share the same population correlation matrix at each grid point, while the marginal signal configuration and weight rule remain fixed. Write $x=s_J(\Gamma)/\Delta$. Cert-wBY denotes the proposed rule with a weighted BY fallback, and direct wBY applies weighted BY at $q$ without certification. The population-sign selector uses the weighted BH branch when all checked population correlations are nonnegative and the fallback otherwise. Diagnostic wBH applies weighted BH at $q$ without certification. The values \(x=-0.25\), \(0.50\), and \(1\) represent the negative-sign regime, the transition region, and the sufficient completeness threshold, respectively. At \(x=-0.25\), Cert-wBY had \(0.0131\) lower mean power than direct wBY, with 95\% HB interval \([-0.0214,-0.0048]\). At \(x=0.50\), certificate acceptance was \(0.2201\), and the mean-power difference became \(0.0387\), with interval \([0.0304,0.0470]\). At \(x=1\), the difference was \(0.2229\), with interval \([0.2145,0.2312]\).

\begin{figure}[!htb]
\centering
\includegraphics[width=0.95\linewidth]
{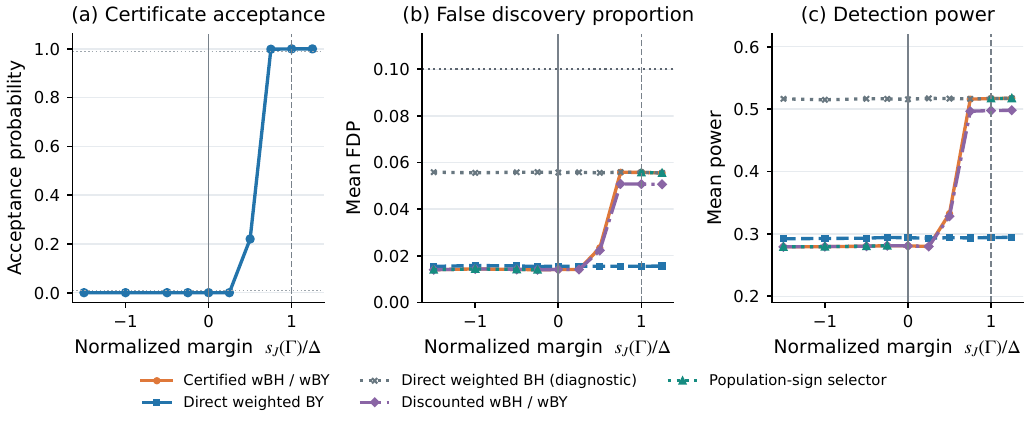}
\caption{Routing along the Gaussian dependence path $x=s_J(\Gamma)/\Delta$. Panels (a)--(c) show certificate acceptance, mean FDP, and mean power. Vertical lines mark $x=0$ and $x=1$; horizontal lines in (a) mark $\beta$ and $1-\beta$. The population-sign selector is shown only for $x<0$ or $x\ge1$. Uncertified wBH is diagnostic. Monte Carlo intervals appear in \cref{app:routing-transition}.}
\label{fig:routing-transition}
\end{figure}

The comparator that discounts both branches uses $q^\circ$ in the certified branch as well as the fallback. Its comparison with Cert-wBY isolates the gain from using $q$ after certification. At
\(x=0.50\), the mean-power difference was \(0.0043\), with interval
\([-0.0040,0.0127]\), and at \(x=1\) it was \(0.0193\), with interval
\([0.0110,0.0276]\). Across the transition grid, the largest upper endpoint
of the prespecified 95\% HB intervals for the mean FDP of Cert-wBY was
\(0.0603<q=0.10\), which is consistent with the finite-sample FDR guarantee
in \cref{thm:sign-certificate}. Diagnostic weighted BH also remained below
\(q\) in these simulated laws, so the experiment does not demonstrate an
empirical FDR failure of weighted BH.

\paragraph{Behavior inside the transition.}
The original grid has only two points in the interval where acceptance changes sharply. We therefore evaluated the refined grid $x=0.25,0.30,\ldots,0.75$, with 50,000 additional Gaussian replications at each point. We kept $m=20$, $N_C=5000$, $q=0.10$, $\beta=0.01$, and the original weights and nonnull configuration. The testing and learning samples within each replication were reused across the compared procedures.

Certificate acceptance increased from $0.0520$ at $x=0.45$ to $0.5191$ at $x=0.55$ and $0.9465$ at $x=0.65$. At $x=0.55$, the mean-power difference between Cert-wBY and direct wBY was $0.1108$, with a simultaneous 95\% HB interval of $[0.0952,0.1264]$. Across the 11 points, the maximum mean FDP of Cert-wBY was $0.0558$, with a simultaneous upper bound of $0.0636$.

\begin{figure}[!tbp]
\centering
\includegraphics[width=.93\linewidth]{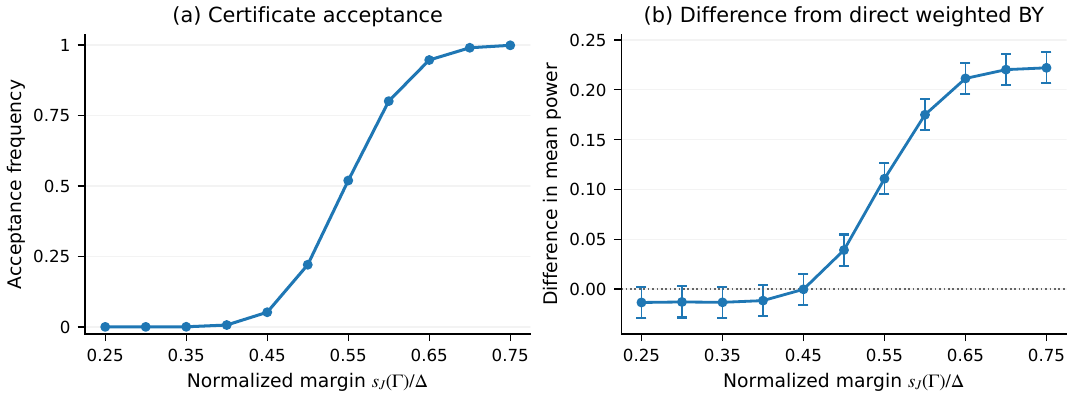}
\caption{Certificate acceptance (a) and paired mean-power difference between Cert-wBY and direct wBY (b) on the refined margin grid. Error bars in (b) are simultaneous 95\% HB intervals.}
\label{fig:routing-zoom}
\end{figure}

\FloatBarrier

\paragraph{Certificate calibration and sample size.}
The second experiment isolates the certificate and varies the number of
checked correlations $d\in\{5,20,100\}$ and the population margin
$\gamma\in\{0.05,0.10,0.20\}$. We use the nominal multiplier of the sufficient sample size $r$, so the number of certification scores is
\[
N_C
=
\left\lceil r\left\lceil
\frac{4\log(d/\beta)}{\{\sqrt{1+\gamma}-1\}^2}
\right\rceil\right\rceil.
\]
The original Monte Carlo experiment used
$r\in\{1/8,1/4,1/2,1\}$, with 200,000 replications per
configuration. Across the nine positive-margin configurations,
certificate acceptance was between $0.0181$ and $0.2320$ at $r=1/4$
and between $0.9869$ and $0.9996$ at $r=1/2$. It was one
in every configuration at $r=1$.

We resolve this transition more finely using the exact independent-block
distribution from \cref{app:certificate-calibration}. For each of the nine
$(d,\gamma)$ settings, we evaluate acceptance by one-dimensional
numerical integration at 11 multipliers between $0.25$ and $0.50$.
At $r=0.30$, the nine acceptance probabilities range from $0.4089$
to $0.5144$. At $r=0.40$, they range from $0.8952$ to $0.9774$.
The Monte Carlo estimates at $r=0.25$ and $r=0.50$ are consistent with the numerical integration results. The intermediate values are numerical integration
results rather than additional Monte Carlo replications.

Across the original full grid, the certificate accepted none of
7.2 million negative-sign replications and 18 of 7.2 million
boundary replications. At the sufficient bound, it accepted all
1.8 million positive-margin replications. These results illustrate
the finite-sample sufficient bound and its conservative sample-size
constant. They do not estimate the minimax constant in
\cref{thm:certification-lower,thm:certification-lower-all-pairs}.

\begin{figure}[!htb]
\centering
\includegraphics[width=\textwidth]
{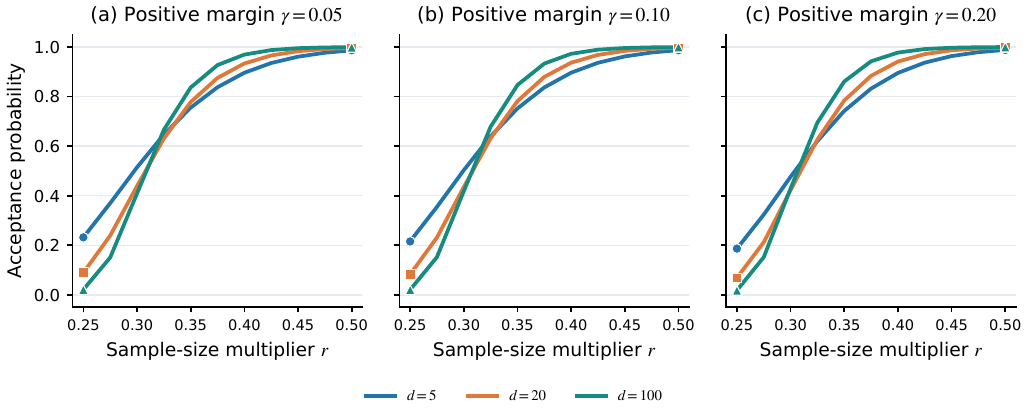}
\caption{Certificate acceptance under positive-margin Gaussian block models. Panels fix $\gamma=0.05$, $0.10$, and $0.20$; curves distinguish $d=5$, $20$, and $100$. Solid curves are numerical evaluations of \eqref{eq:block-acceptance}; endpoint symbols are Monte Carlo estimates.}
\label{fig:certificate-calibration}
\end{figure}

\paragraph{Moderate-dimensional stress test.}
We next repeat the end-to-end routing comparison with
\(m\in\{20,50,100\}\) and the all-pairs choice of \(J\), giving
\(d\in\{190,1225,4950\}\). The certification sample size is adjusted to
keep the completeness margin essentially matched across dimensions. The
same three normalized-margin regimes are used at every dimension. This
experiment therefore changes the multiplicity of the all-pairs problem
without introducing a new routing mechanism.

In the simulated replications, the certificate never accepted at $x=-0.5$ and always accepted at $x=1.25$ for all three dimensions. At $x=0.5$, the acceptance frequency decreased as $m$ increased. The 95\% HB upper endpoints for the mean FDP of Cert-wBY
remained below \(q=0.10\). The sign of the power contrast with direct wBY
was negative at \(x=-0.5\) and
positive at \(x=0.5\) and \(x=1.25\). The design is described in
\cref{app:moderate-dimension-stress}.

\begin{figure}[!tbp]
\centering
\includegraphics[width=\textwidth]
{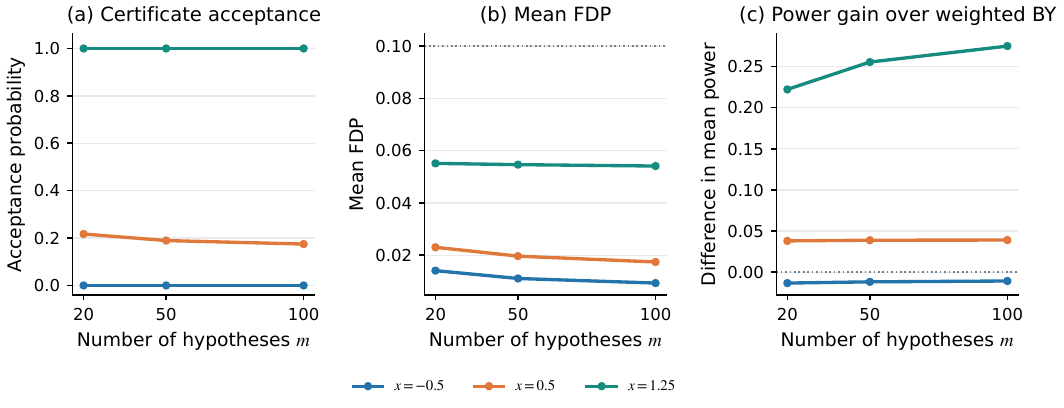}
\caption{Certificate acceptance (a), mean FDP for Cert-wBY (b), and paired mean-power difference relative to direct wBY (c) across dimensions. Dotted lines in (b) and (c) mark $q$ and zero. See \cref{app:moderate-dimension-stress} for the design and intervals.}
\label{fig:moderate-dimension-stress}
\end{figure}

\paragraph{Certification and testing under a fixed subject budget.}
The Gaussian construction in \cref{prop:gaussian-same-subject} allows us to compare procedures that use the same subjects and separately learned weights. We simulated six Gaussian correlation structures with $m=20$ hypotheses, four nonnull means equal to $2.5/\sqrt{600}$ at the subject level, and 20,000 replications per regime. Each method used 600 study subjects and a common independent pilot of 100 additional subjects to learn the weights. The reuse procedure tested with the mean of all 600 study subjects and certified with 599 orthogonal contrasts. A split comparator tested with 300 subjects and certified using 150 paired differences from the remaining 300. Direct weighted BY used all 600 subjects for testing. The split comparison combines differences in subject allocation and reference-score construction; see \cref{app:fixed-budget-study} for details.

Under equicorrelation $0.30$, the reused-subject procedure had mean power $0.5954$ versus $0.4668$ for direct weighted BY. Under heterogeneous positive dependence, the corresponding powers were $0.6773$ and $0.4726$. The paired differences were $0.1286$ and $0.2046$, exceeding the simultaneous 95\% half-width of $0.0234$ for the six regimes. No certifications occurred under mixed signs, independence, or weak positive dependence in the 20,000 replications per regime. The associated power losses relative to direct weighted BY ranged from $0.0143$ to $0.0150$. The largest mean FDP of the certified reuse procedure was $0.0585$, with a simultaneous 95\% upper bound of $0.0702$.

\begin{figure}[!htbp]
\centering
\includegraphics[width=.98\linewidth]{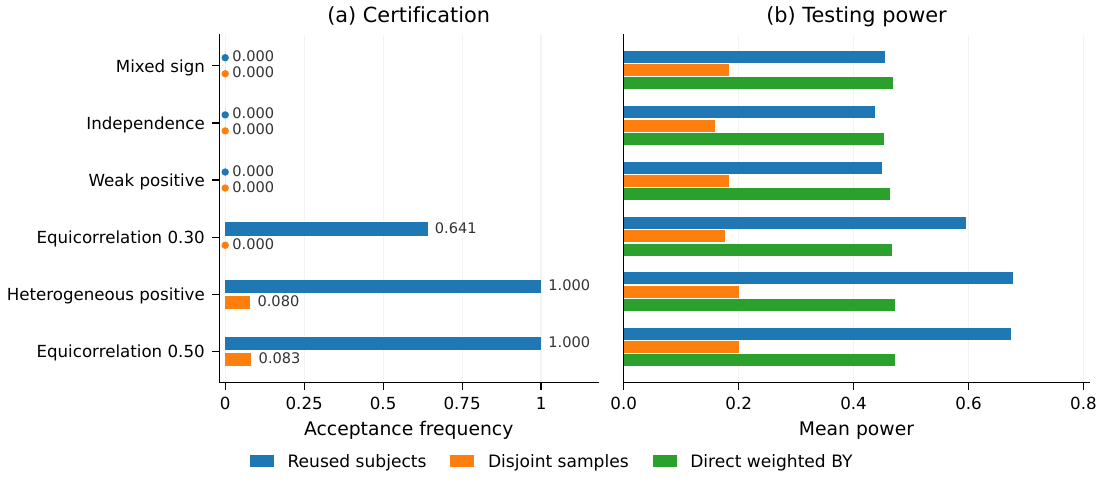}
\caption{Certificate acceptance for the two certified procedures (a) and mean power for all three procedures (b) under a common subject budget. See \cref{tab:fixed-budget-comparison}.}
\label{fig:fixed-budget}
\end{figure}

\FloatBarrier

\paragraph{Gaussian simulation using HBN correlations.}
We estimate a $10\times10$ correlation matrix from the HBN learning split and treat it as the population matrix of a new Gaussian simulation. The primary setting uses signal strength $\nu=2$, weight parameter $\eta=1$, and $N_C=682$ reference scores, as defined in \cref{app:hbn-bridge}. Here $s_J(\Gamma)/\Delta=0.5611$. Certificate acceptance was $0.9358$, with 95\% CP interval $[0.9343,0.9374]$. Although the minimum correlation is below the sufficient margin, acceptance is frequent in this particular geometry.

Cert-wBY attained mean power $0.5445$, compared with $0.3734$ for direct wBY. Its mean FDP was $0.0480$, with a 95\% HB upper endpoint of $0.0542<q$. \Cref{tab:module-c-bridge} compares the methods at this setting. The full simulation design and HBN provenance appear in \cref{app:hbn-bridge,app:hbn-provenance}.

\begin{table}[!htbp]
\centering
\caption{Gaussian simulation using HBN correlations at the primary setting. Brackets give simultaneous 95\% HB intervals over 63 prespecified means for each metric. Weighted BH is diagnostic.}
\label{tab:module-c-bridge}
\begingroup
\small
\setlength{\tabcolsep}{4pt}
\renewcommand{\arraystretch}{1.08}
\begin{tabular*}{\linewidth}{@{\extracolsep{\fill}}lcc@{}}
\toprule
Method & Mean FDP [95\% HB] & Mean power [95\% HB] \\
\midrule
Cert-wBY (proposed) & 0.0480 [0.0417, 0.0542] & 0.5445 [0.5382, 0.5507] \\
two-discounted branch & 0.0436 [0.0374, 0.0499] & 0.5280 [0.5218, 0.5343] \\
direct wBY($q$) & 0.0174 [0.0112, 0.0237] & 0.3734 [0.3671, 0.3796] \\
diagnostic wBH($q$) & 0.0503 [0.0440, 0.0566] & 0.5572 [0.5509, 0.5634] \\
\bottomrule
\end{tabular*}

\endgroup
\end{table}

The separate analysis of the observed HBN participants yielded uniform weights and ten rejections under every reported method. It documents the implementation but cannot distinguish the procedures in this setting. We report its assumptions and coverage limitations in \cref{app:hbn-provenance}.
\FloatBarrier

\section{Discussion}

Reference data can support the use of a more powerful weighted testing rule without requiring its dependence condition to be assumed in advance. The boundary obstruction and minimax bounds explain why uniform certification becomes difficult near zero correlation. Under the Gaussian model, the routed procedure controls FDR even when the sign condition fails. A more powerful fallback could improve detection if it satisfies the conditional bound in \cref{thm:adjusted-certified-fdr}.

The sign condition is sufficient for weighted BH but does not characterize all Gaussian laws under which that rule is valid. In particular, some negative correlations are compatible with FDR control. Diagnostic weighted BH remained below the target FDR in the six regimes of the fixed-budget experiment. The example in \cref{ex:weighted-bh-inflation} shows that a violation is possible. The numerical comparisons measure the power cost of using the fallback and the gains when certification succeeds. The comparison with Gaussian exponential e-values in \cref{prop:exp-evalue-comparison} is limited to that construction. Comparisons with compound e-values and conditional calibration remain open. The exact threshold in \eqref{eq:exact-threshold} can improve acceptance without changing the routing guarantee. Its numerical comparison with the explicit concentration threshold is left for future work.

The applicability of the Gaussian certificate depends on how the reference and testing scores are obtained. The construction in \cref{prop:gaussian-same-subject} is useful for multiple endpoints measured on independent subjects with a common Gaussian distribution and known marginal variances. It avoids reserving subjects exclusively for certification. Site differences, family dependence, and unknown marginal scales may invalidate the assumptions on the scores. In the HBN analysis, variances were estimated in the learning sample and family identifiers were unavailable. We therefore have not established conditional normality or equality of the population correlation matrices. Nor do the observed data provide known null labels. The Gaussian simulations quantify power and FDR under their specified models. The observed-data analysis illustrates preprocessing and implementation rather than verifying finite-sample FDR control. Certification under covariance shift, weaker sufficient sign conditions, and calibration after variance estimation remain open problems.

\paragraph{Code and data availability.}
The numerical designs and preprocessing procedures are described in \cref{app:additional-experiments}. A subsequent reproducibility release is planned to include simulation code, aggregate numerical outputs, and the full-precision HBN-derived correlation matrix, subject to the original data-use conditions. The fixed ROI order is reported, although the original label-ranking procedure cannot be reconstructed exactly. Reproduction from participant-level data is subject to the access and use conditions in \cref{app:hbn-provenance}.

\paragraph{AI use statement.}
The author formulated the research questions, developed the initial mathematical arguments and manuscript drafts, and directed the methodological and numerical studies. ChatGPT (GPT-5.6) was used during the revision process to audit arguments and suggest clarifications, additional derivations, and edits. In preparing this first arXiv version, the author primarily used the GPT-6 series of ChatGPT to check the mathematical exposition and presentation. AI assistance also included suggestions for supplementary calculations and simulations, code implementation, and figure preparation. The author evaluated the suggestions, decided which to retain, modify, or reject, and determined the scope and interpretation of the research. The author takes responsibility for the mathematical statements, proofs, numerical results, and conclusions. These checks supplement the author's evaluation and do not constitute independent peer review.

\bibliographystyle{apalike}
\bibliography{bibliography}

\clearpage
\appendix
\section{Additional numerical results}
\label{app:additional-experiments}

\subsection{Gaussian routing transition}
\label{app:routing-transition}

The transition experiment used \(m=20\),
$J=\{(i,j):1\le i<j\le20\}$, $d=190$, and $N_C=5000$.
For normalized margin
\[
x=s_J(\Gamma)/\Delta
\in
\{-1.50,-1.00,-0.50,-0.25,0,0.25,0.50,0.75,1,1.25\},
\]
we set
\[
\Gamma(x)
=
a(x)a(x)^\top
+
\operatorname{diag}\{1-a_i(x)^2\},
\]
where
\[
a_i(x)=0.5,\quad i\le19,
\qquad
a_{20}(x)=2x\Delta.
\]
Hence \(s_J\{\Gamma(x)\}=x\Delta\). Across the specified grid,
\(|a_{20}(x)|<1\), so every \(\Gamma(x)\) is a correlation matrix.

The four nonnull coordinates were $\{1,2,11,12\}$, with $\mu_i=2$ for these coordinates and $\mu_i=0$ elsewhere. For each $x$, the learning score $Z^W$ and testing score $Z$ were independent draws from $N_m(\mu,\Gamma(x))$.
For $a\le b$, define $\operatorname{clip}(u,a,b)=\min\{b,\max\{a,u\}\}$. For the learning score \(Z^W\), the normalized weights were
\begin{equation}\label{eq:routing-weights}
W_i
=
\frac{
m\,\operatorname{clip}\{\exp(Z_i^W),0.25,4\}
}{
\sum_{j=1}^m
\operatorname{clip}\{\exp(Z_j^W),0.25,4\}
},
\end{equation}
corresponding to the prespecified weight parameter \(\eta=1\).
Independently of the learning and testing scores, the certification
cross-moment matrix satisfied
\[
N_C\widehat\Gamma
\sim
\operatorname{Wishart}_m\{N_C,\Gamma(x)\}.
\]
The three information sources therefore shared the same population
correlation matrix at each grid point, while the common dependence law
changed across the grid. The full testing distribution was not held fixed
while the certificate alone was varied.

Each grid point used 200,000 replications. In each replication the same weights and testing scores were used for the seven procedures. Direct BY and direct wBY apply $\wBH(\mathbf1_m,q/c_m)$ and $\wBH(W,q/c_m)$, respectively. Diagnostic wBH applies $\wBH(W,q)$ without certification. Cert-BY and Cert-wBY use the prescribed BY and weighted BY fallbacks. The population-sign selector $R^\star$ uses the population sign as defined in \cref{cor:branch-selector-comparison}, with $R^{\mathrm{wBY}}$ as its fallback. The discounted wBH/wBY comparator applies weighted BH at $q^\circ$ upon certification and the same weighted BY fallback otherwise. For FDP and power separately,
each HB family contains 70 prespecified means from the ten grid points and
seven procedures. The 26 paired power contrasts form a separate HB family
and use the \([-1,1]\) range of a paired difference. Figure
\ref{fig:routing-transition} shows the transition curves, and the tables
below report the full numerical results and paired contrasts.

\begingroup
\normalsize
\setlength{\tabcolsep}{6pt}
\renewcommand{\arraystretch}{1.15}
\begingroup
\begin{longtable}{@{}rlcc@{}}
\caption{
Full results for the Gaussian routing transition. HB denotes the Hoeffding--Bonferroni interval for a bounded mean.
Daggered values for the population-sign branch selector in \(0\le s_J/\Delta<1\) are descriptive because the comparison guarantee in \cref{cor:branch-selector-comparison} does not cover that region.
}
\label{tab:supplemental-routing-methods} \\
\toprule
$s_J/\Delta$ & Method & Mean FDP [95\% HB] & Mean power [95\% HB] \\
\midrule
\endfirsthead
\multicolumn{4}{@{}l}{\tablename~\thetable\ (continued)} \\
\toprule
$s_J/\Delta$ & Method & Mean FDP [95\% HB] & Mean power [95\% HB] \\
\midrule
\endhead
-1.50 & BY($q$) & 0.0206 [0.0162, 0.0251] & 0.2125 [0.2080, 0.2170] \\
-1.50 & wBY($q$) & 0.0154 [0.0110, 0.0199] & 0.2921 [0.2877, 0.2966] \\
-1.50 & wBH($q$), diagnostic & 0.0559 [0.0514, 0.0603] & 0.5164 [0.5120, 0.5209] \\
-1.50 & Cert-wBY & 0.0140 [0.0096, 0.0185] & 0.2791 [0.2747, 0.2836] \\
-1.50 & Cert-BY & 0.0187 [0.0142, 0.0231] & 0.2020 [0.1976, 0.2065] \\
-1.50 & population-sign branch selector & 0.0140 [0.0096, 0.0185] & 0.2791 [0.2747, 0.2836] \\
-1.50 & two-discounted branch & 0.0140 [0.0096, 0.0185] & 0.2791 [0.2747, 0.2836] \\
-1.00 & BY($q$) & 0.0207 [0.0162, 0.0251] & 0.2120 [0.2076, 0.2165] \\
-1.00 & wBY($q$) & 0.0158 [0.0114, 0.0203] & 0.2925 [0.2881, 0.2970] \\
-1.00 & wBH($q$), diagnostic & 0.0556 [0.0511, 0.0600] & 0.5149 [0.5105, 0.5194] \\
-1.00 & Cert-wBY & 0.0144 [0.0099, 0.0188] & 0.2794 [0.2750, 0.2839] \\
-1.00 & Cert-BY & 0.0187 [0.0143, 0.0232] & 0.2017 [0.1972, 0.2061] \\
-1.00 & population-sign branch selector & 0.0144 [0.0099, 0.0188] & 0.2794 [0.2750, 0.2839] \\
-1.00 & two-discounted branch & 0.0144 [0.0099, 0.0188] & 0.2794 [0.2750, 0.2839] \\
-0.50 & BY($q$) & 0.0209 [0.0164, 0.0253] & 0.2129 [0.2084, 0.2173] \\
-0.50 & wBY($q$) & 0.0156 [0.0112, 0.0201] & 0.2931 [0.2886, 0.2975] \\
-0.50 & wBH($q$), diagnostic & 0.0558 [0.0514, 0.0603] & 0.5165 [0.5121, 0.5210] \\
-0.50 & Cert-wBY & 0.0142 [0.0097, 0.0186] & 0.2801 [0.2756, 0.2845] \\
-0.50 & Cert-BY & 0.0190 [0.0145, 0.0234] & 0.2026 [0.1981, 0.2070] \\
-0.50 & population-sign branch selector & 0.0142 [0.0097, 0.0186] & 0.2801 [0.2756, 0.2845] \\
-0.50 & two-discounted branch & 0.0142 [0.0097, 0.0186] & 0.2801 [0.2756, 0.2845] \\
-0.25 & BY($q$) & 0.0204 [0.0160, 0.0249] & 0.2140 [0.2095, 0.2184] \\
-0.25 & wBY($q$) & 0.0154 [0.0109, 0.0198] & 0.2941 [0.2897, 0.2986] \\
-0.25 & wBH($q$), diagnostic & 0.0559 [0.0514, 0.0603] & 0.5164 [0.5120, 0.5209] \\
-0.25 & Cert-wBY & 0.0140 [0.0096, 0.0185] & 0.2811 [0.2766, 0.2855] \\
-0.25 & Cert-BY & 0.0186 [0.0141, 0.0230] & 0.2037 [0.1993, 0.2082] \\
-0.25 & population-sign branch selector & 0.0140 [0.0096, 0.0185] & 0.2811 [0.2766, 0.2855] \\
-0.25 & two-discounted branch & 0.0140 [0.0096, 0.0185] & 0.2811 [0.2766, 0.2855] \\
0.00 & BY($q$) & 0.0206 [0.0161, 0.0250] & 0.2131 [0.2086, 0.2176] \\
0.00 & wBY($q$) & 0.0155 [0.0111, 0.0200] & 0.2937 [0.2893, 0.2982] \\
0.00 & wBH($q$), diagnostic & 0.0557 [0.0513, 0.0602] & 0.5157 [0.5113, 0.5202] \\
0.00 & Cert-wBY & 0.0141 [0.0097, 0.0186] & 0.2807 [0.2762, 0.2851] \\
0.00 & Cert-BY & 0.0186 [0.0142, 0.0231] & 0.2028 [0.1984, 0.2073] \\
0.00 & population-sign branch selector$^\dagger$ & 0.0557 [0.0513, 0.0602] & 0.5157 [0.5113, 0.5202] \\
0.00 & two-discounted branch & 0.0141 [0.0097, 0.0186] & 0.2807 [0.2762, 0.2851] \\
0.25 & BY($q$) & 0.0205 [0.0161, 0.0250] & 0.2127 [0.2082, 0.2171] \\
0.25 & wBY($q$) & 0.0155 [0.0111, 0.0200] & 0.2930 [0.2886, 0.2975] \\
0.25 & wBH($q$), diagnostic & 0.0558 [0.0513, 0.0602] & 0.5172 [0.5128, 0.5217] \\
0.25 & Cert-wBY & 0.0141 [0.0097, 0.0186] & 0.2798 [0.2753, 0.2842] \\
0.25 & Cert-BY & 0.0186 [0.0142, 0.0231] & 0.2026 [0.1981, 0.2070] \\
0.25 & population-sign branch selector$^\dagger$ & 0.0558 [0.0513, 0.0602] & 0.5172 [0.5128, 0.5217] \\
0.25 & two-discounted branch & 0.0141 [0.0097, 0.0186] & 0.2798 [0.2753, 0.2842] \\
0.50 & BY($q$) & 0.0207 [0.0162, 0.0251] & 0.2135 [0.2090, 0.2179] \\
0.50 & wBY($q$) & 0.0155 [0.0111, 0.0200] & 0.2938 [0.2893, 0.2982] \\
0.50 & wBH($q$), diagnostic & 0.0556 [0.0511, 0.0600] & 0.5169 [0.5125, 0.5214] \\
0.50 & Cert-wBY & 0.0234 [0.0190, 0.0279] & 0.3325 [0.3280, 0.3369] \\
0.50 & Cert-BY & 0.0271 [0.0227, 0.0316] & 0.2722 [0.2678, 0.2767] \\
0.50 & population-sign branch selector$^\dagger$ & 0.0556 [0.0511, 0.0600] & 0.5169 [0.5125, 0.5214] \\
0.50 & two-discounted branch & 0.0223 [0.0178, 0.0267] & 0.3282 [0.3237, 0.3326] \\
0.75 & BY($q$) & 0.0209 [0.0165, 0.0254] & 0.2133 [0.2088, 0.2177] \\
0.75 & wBY($q$) & 0.0155 [0.0111, 0.0200] & 0.2935 [0.2890, 0.2979] \\
0.75 & wBH($q$), diagnostic & 0.0559 [0.0514, 0.0603] & 0.5164 [0.5119, 0.5208] \\
0.75 & Cert-wBY & 0.0558 [0.0513, 0.0603] & 0.5160 [0.5115, 0.5204] \\
0.75 & Cert-BY & 0.0558 [0.0514, 0.0603] & 0.5159 [0.5114, 0.5203] \\
0.75 & population-sign branch selector$^\dagger$ & 0.0559 [0.0514, 0.0603] & 0.5164 [0.5119, 0.5208] \\
0.75 & two-discounted branch & 0.0508 [0.0464, 0.0553] & 0.4965 [0.4921, 0.5010] \\
1.00 & BY($q$) & 0.0210 [0.0166, 0.0255] & 0.2133 [0.2088, 0.2178] \\
1.00 & wBY($q$) & 0.0155 [0.0110, 0.0200] & 0.2941 [0.2896, 0.2985] \\
1.00 & wBH($q$), diagnostic & 0.0558 [0.0513, 0.0602] & 0.5169 [0.5125, 0.5214] \\
1.00 & Cert-wBY & 0.0558 [0.0513, 0.0602] & 0.5169 [0.5125, 0.5214] \\
1.00 & Cert-BY & 0.0558 [0.0513, 0.0602] & 0.5169 [0.5125, 0.5214] \\
1.00 & population-sign branch selector & 0.0558 [0.0513, 0.0602] & 0.5169 [0.5125, 0.5214] \\
1.00 & two-discounted branch & 0.0508 [0.0463, 0.0552] & 0.4976 [0.4932, 0.5021] \\
1.25 & BY($q$) & 0.0209 [0.0165, 0.0254] & 0.2134 [0.2090, 0.2179] \\
1.25 & wBY($q$) & 0.0156 [0.0111, 0.0201] & 0.2944 [0.2899, 0.2989] \\
1.25 & wBH($q$), diagnostic & 0.0555 [0.0510, 0.0600] & 0.5173 [0.5128, 0.5217] \\
1.25 & Cert-wBY & 0.0555 [0.0510, 0.0600] & 0.5173 [0.5128, 0.5217] \\
1.25 & Cert-BY & 0.0555 [0.0510, 0.0600] & 0.5173 [0.5128, 0.5217] \\
1.25 & population-sign branch selector & 0.0555 [0.0510, 0.0600] & 0.5173 [0.5128, 0.5217] \\
1.25 & two-discounted branch & 0.0507 [0.0462, 0.0551] & 0.4981 [0.4937, 0.5026] \\
\bottomrule
\end{longtable}
\endgroup

\endgroup

\medskip
\begingroup
\normalsize
\setlength{\tabcolsep}{6pt}
\renewcommand{\arraystretch}{1.15}
\begin{longtable}{@{}rllc@{}}
\caption{
Prespecified differences in mean power, computed as Cert-wBY minus the
listed comparator. Contrasts with the population-sign branch selector are reported only where the comparison guarantee in \cref{cor:branch-selector-comparison} applies.
}
\label{tab:supplemental-routing-contrasts} \\
\toprule
$s_J/\Delta$ & Left method & Right method & Mean power difference [95\% HB] \\
\midrule
\endfirsthead
\multicolumn{4}{@{}l}{\tablename~\thetable\ (continued)} \\
\toprule
$s_J/\Delta$ & Left method & Right method & Mean power difference [95\% HB] \\
\midrule
\endhead
-1.50 & Cert-wBY & two-discounted branch & 0.0000 [-0.0083, 0.0083] \\
-1.50 & Cert-wBY & population-sign branch selector & 0.0000 [-0.0083, 0.0083] \\
-1.50 & Cert-wBY & wBY($q$) & -0.0130 [-0.0213, -0.0046] \\
-1.00 & Cert-wBY & two-discounted branch & 0.0000 [-0.0083, 0.0083] \\
-1.00 & Cert-wBY & population-sign branch selector & 0.0000 [-0.0083, 0.0083] \\
-1.00 & Cert-wBY & wBY($q$) & -0.0131 [-0.0214, -0.0047] \\
-0.50 & Cert-wBY & two-discounted branch & 0.0000 [-0.0083, 0.0083] \\
-0.50 & Cert-wBY & population-sign branch selector & 0.0000 [-0.0083, 0.0083] \\
-0.50 & Cert-wBY & wBY($q$) & -0.0130 [-0.0213, -0.0046] \\
-0.25 & Cert-wBY & two-discounted branch & 0.0000 [-0.0083, 0.0083] \\
-0.25 & Cert-wBY & population-sign branch selector & 0.0000 [-0.0083, 0.0083] \\
-0.25 & Cert-wBY & wBY($q$) & -0.0131 [-0.0214, -0.0048] \\
0.00 & Cert-wBY & two-discounted branch & 0.0000 [-0.0083, 0.0083] \\
0.00 & Cert-wBY & wBY($q$) & -0.0131 [-0.0214, -0.0047] \\
0.25 & Cert-wBY & two-discounted branch & 0.0000 [-0.0083, 0.0083] \\
0.25 & Cert-wBY & wBY($q$) & -0.0133 [-0.0216, -0.0049] \\
0.50 & Cert-wBY & two-discounted branch & 0.0043 [-0.0040, 0.0127] \\
0.50 & Cert-wBY & wBY($q$) & 0.0387 [0.0304, 0.0470] \\
0.75 & Cert-wBY & two-discounted branch & 0.0195 [0.0111, 0.0278] \\
0.75 & Cert-wBY & wBY($q$) & 0.2225 [0.2142, 0.2309] \\
1.00 & Cert-wBY & two-discounted branch & 0.0193 [0.0110, 0.0276] \\
1.00 & Cert-wBY & population-sign branch selector & 0.0000 [-0.0083, 0.0083] \\
1.00 & Cert-wBY & wBY($q$) & 0.2229 [0.2145, 0.2312] \\
1.25 & Cert-wBY & two-discounted branch & 0.0192 [0.0108, 0.0275] \\
1.25 & Cert-wBY & population-sign branch selector & 0.0000 [-0.0083, 0.0083] \\
1.25 & Cert-wBY & wBY($q$) & 0.2229 [0.2145, 0.2312] \\
\bottomrule
\end{longtable}

\endgroup

\paragraph{Refined grid.}
The experiment in \cref{fig:routing-zoom} uses the same Gaussian model, weights in \eqref{eq:routing-weights}, and seven procedures on $x\in\{0.25,0.30,\ldots,0.75\}$. Each of the 11 configurations uses 50,000 independent replications. The paired mean-power differences between Cert-wBY and direct wBY form one HB family of size 11. The mean FDPs of Cert-wBY form a separate family of size 11; both use \eqref{eq:hb-halfwidth}.

\FloatBarrier

\subsection{Certificate-calibration design}
\label{app:certificate-calibration}

The certificate-calibration experiment used
$d\in\{5,20,100\}$, $\gamma\in\{0.05,0.10,0.20\}$, and nominal multipliers $r\in\{1/8,1/4,1/2,1\}$.
For each \((d,\gamma,r)\), the certification sample size was
\[
N_C
=
\left\lceil
r
\left\lceil
\frac{4\log(d/\beta)}
     {\{\sqrt{1+\gamma}-1\}^2}
\right\rceil
\right\rceil .
\]
The inner ceiling rounds the sufficient bound in \eqref{eq:sufficient-size}. Because $N_C$ is rounded twice, \(r\) is a nominal multiplier of the
sufficient bound and need not equal the realized ratio of \(N_C\) to that
bound.

The certification law consisted of $d$ independent Gaussian $2\times2$ blocks, with $m=2d$ and $J$ containing the within-block pairs. For a block with population correlation $\rho$, the empirical cross-moment was sampled exactly according to \eqref{eq:cross-moment-law}.
In the positive-margin configuration every certified correlation was
\(\gamma\). The boundary and negative-sign configurations replaced the
first certified correlation by $0$ and $-\gamma$, respectively, while keeping the remaining $d-1$ correlations at $+\gamma$.

The design contains \(3\times3\times4\times3=108\) configurations,
each evaluated with 200,000 replications. Of the 27 configurations
with \(r=1\), nine correspond to the positive-margin law.
In \cref{fig:certificate-calibration}, the solid curves supplement these results with numerical integration of the exact distribution at
\(r\in\{0.25,0.275,\ldots,0.50\}\). For independent Gaussian pairs,
\begin{equation}\label{eq:block-acceptance}
\Pr(C=1)
=
\left[
\int_{0}^{\infty}
F_{\chi^2_{N_C}}
\!\left(\frac{(1+\gamma)x-2N_C\tau}{1-\gamma}\right)
 f_{\chi^2_{N_C}}(x)\,dx
\right]^d,
\end{equation}
where \(F_{\chi^2_{N_C}}\) and \(f_{\chi^2_{N_C}}\) denote the CDF and
density of a chi-square random variable with \(N_C\) degrees of
freedom. The integral follows from the displayed cross-moment
representation and independence across the \(d\) Gaussian blocks.
 The table below reports aggregate
acceptance counts from the original Monte Carlo experiment.

\begin{table}[!tbp]
\centering
\caption{Aggregate certificate-calibration counts.}
\label{tab:certificate-aggregate}
\begin{tabular*}{0.88\linewidth}{@{\extracolsep{\fill}}lrr@{}}
\toprule
Configuration & Replications & Acceptances \\
\midrule
Negative sign & 7,200,000 & 0 \\
Boundary & 7,200,000 & 18 \\
Positive margin at the sufficient bound
& 1,800,000 & 1,800,000 \\
\bottomrule
\end{tabular*}
\end{table}

\subsection{Moderate-dimensional routing stress test}
\label{app:moderate-dimension-stress}

We repeated the Gaussian routing experiment for $m=20,50,100$. The all-pairs choice of $J$ gives $d=190,1225,4950$, respectively. The normalized margins were $x=s_J(\Gamma)/\Delta\in\{-0.5,0.5,1.25\}$.
We retained the one-factor construction used in the routing-transition
experiment. For each \(m\), its loading vector was
\[
a_i(x)=0.5,
\qquad i=1,\ldots,m-1,
\qquad
a_m(x)=2x\Delta,
\]
where \(\Delta\) was evaluated using the corresponding values of
\(d\), \(N_C\), and \(\beta\).
The resulting correlation matrix satisfies
\[
s_J\{\Gamma(x)\}=x\Delta
\]
for the all-pairs choice of \(J\).
The score-scale signal strength, weight rule \eqref{eq:routing-weights}, and testing procedures were otherwise unchanged.

The certification sample sizes were \(5000\), \(5559\), and \(5967\) for
\(m=20\), \(50\), and \(100\), respectively. These choices approximately
match the completeness margin of the original \(m=20\) design after
integer rounding. The nonnull proportion remained \(0.20\). The nonnull
coordinates were \(\{1,2,11,12\}\) for \(m=20\),
\(\{1,\ldots,5,26,\ldots,30\}\) for \(m=50\), and
\(\{1,\ldots,10,51,\ldots,60\}\) for \(m=100\).

Each of the nine configurations used 200,000 replications. For FDP and
power separately, each HB family contained 63 prespecified means from the
nine configurations and seven procedures. The nine paired differences between Cert-wBY and direct wBY formed a separate HB family. \Cref{fig:moderate-dimension-stress} shows the principal results. This study
changes the all-pairs multiplicity while retaining the same Gaussian
routing construction. It is not an empirical test of the minimax lower
bound and does not establish performance on observed connectivity data.

\subsection{Gaussian simulation using HBN correlations}
\label{app:hbn-bridge}

The simulation uses a ten-dimensional correlation
matrix estimated only from the HBN learning split.
A rule based only on region-of-interest (ROI) labels selected the ten coordinates before analysis.
\Cref{app:hbn-provenance} describes the rule
and the fixed coordinate order.

\begin{figure}[!tbp]
\centering
\includegraphics[width=.56\linewidth]{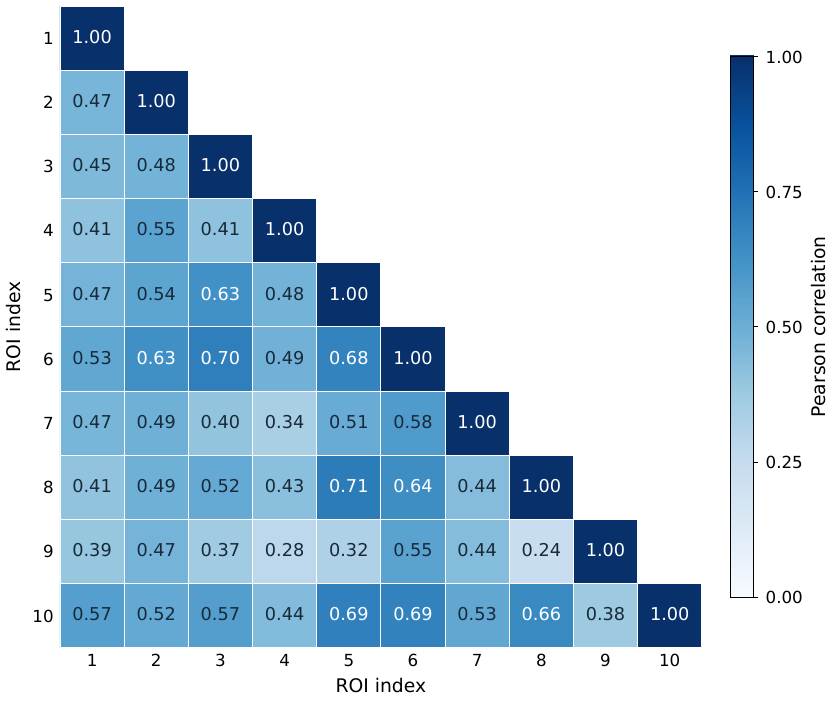}
\caption{Population correlation matrix used in the HBN-derived Gaussian simulations. Coordinates follow the order specified in \cref{app:hbn-provenance}.}
\label{fig:hbn-correlation}
\end{figure}

The archived geometry record preserves the Pearson correlation matrix of the selected learning features after the nuisance adjustment in \eqref{eq:hbn-adjustment}. This fixed matrix, denoted by $\Gamma_{\mathrm{HBN}}$, is the population correlation matrix $\Gamma$ in the Gaussian simulations. It was positive definite without projection, and no nonnegativity constraint was imposed on the off-diagonal entries. As shown in \cref{fig:hbn-correlation}, the off-diagonal correlations range from $0.2383$ to $0.7071$, with minimum eigenvalue $0.2081$.

The first two coordinates were nonnull.
For signal level $\nu\in\{1.5,2,2.5\}$, set $\mu=(\nu,\nu,0,\ldots,0)^\top$.
Conditional on the frozen correlation matrix
\(\Gamma\), the learning and testing score
vectors were generated independently as
\[
Z^W,Z\sim N_{10}(\mu,\Gamma).
\]
Independently, the certification cross-moment
matrix was generated from
\[
N_C\widehat\Gamma
\sim\operatorname{Wishart}_{10}(N_C,\Gamma),
\qquad N_C=682.
\]

The learned weights were
\begin{equation}\label{eq:hbn-weights}
W_i(\eta)
=
\frac{
10\,\operatorname{clip}
\{\exp(\eta Z_i^W),0.25,4\}
}{
\sum_{j=1}^{10}
\operatorname{clip}
\{\exp(\eta Z_j^W),0.25,4\}
},
\qquad
\eta\in\{0,0.5,1\}.
\end{equation}
For a fixed signal level and Monte Carlo
replication, the same pair \((Z^W,Z)\)
was reused across the three values of \(\eta\).
Thus comparisons across \(\eta\) are paired,
while \(Z^W\) and \(Z\) remain independent
within each replication.

Each of the nine \((\nu,\eta)\) configurations used 100,000 replications. The primary configuration was \((2,1)\).

We compared the same seven procedures as in
the Gaussian routing simulations, namely
direct BY, direct wBY, diagnostic weighted BH,
Cert-wBY, Cert-BY, the population-sign branch
selector, and the discounted wBH/wBY comparator.

For FDP and power separately, the HB family
fixed before simulation contained 63 means
from the nine \((\nu,\eta)\) configurations
and seven procedures.

The primary setting has $0<s_J(\Gamma)/\Delta=0.5611<1$, outside the comparison region of \cref{cor:branch-selector-comparison}. The four primary method summaries are in \cref{tab:module-c-bridge}. 

\FloatBarrier

\subsection{Additional Gaussian comparisons with fixed subject budgets}
\label{app:fixed-budget-study}

We use the homogeneous Gaussian model of \cref{prop:gaussian-same-subject} with $m=20$, $n=600$ study subjects and an independent 100-subject learning pilot. The subject mean vector $\theta$ has entries $2.5/\sqrt{600}$ at coordinates $1,2,11,12$ and zero elsewhere. All procedures share the pilot. Its score has law $N_m(\sqrt{100}\,\theta,\Gamma)$, and its weights follow \eqref{eq:routing-weights}. The six covariance regimes each use 20,000 independent replications. Three are independence and equicorrelations $.30$ and $.50$. The other three have $\Gamma=aa^\top+\operatorname{diag}(1-a_i^2)$, with common loading on the first 19 coordinates and the last loading given by $(.7,-.43)$, $(.7,.10)$, and $(.75,.55)$ for the mixed, weak positive, and heterogeneous positive cases, respectively.

The comparison uses two independent group means $\bar V_1,\bar V_2\sim N_m(\theta,\Gamma/300)$, each based on 300 subjects. Let $S_1,S_2$ be the corresponding centered scatters before division by the degrees of freedom. To couple the split and reuse procedures, decompose the second group scatter into $S_2=S_p+S_r$, where $S_p$ is the scatter of the 150 paired contrasts and $S_r$ is the scatter of the remaining 149 orthogonal residual contrasts. Independently of the group means and the pilot, generate mutually independent scatters
\[
S_1\sim\operatorname{Wishart}_m(299,\Gamma),\qquad
S_p\sim\operatorname{Wishart}_m(150,\Gamma),\qquad
S_r\sim\operatorname{Wishart}_m(149,\Gamma).
\]
The full mean and centered scatter are
\begin{equation}\label{eq:full-scatter}
\begin{aligned}
\bar V&=(\bar V_1+\bar V_2)/2,\quad
S=S_1+S_2+150(\bar V_1-\bar V_2)(\bar V_1-\bar V_2)^\top.
\end{aligned}
\end{equation}
The reuse procedure uses the testing score $\sqrt{600}\,\bar V$ and the certification matrix $S/599$ from \eqref{eq:full-scatter}. The split procedure uses $\sqrt{300}\,\bar V_1$ and $S_p/150$. The sum and difference of the independent group means are independent Gaussian vectors. Hence the full testing score is independent of its certification scatter, which has 599 degrees of freedom. The split score is independent of $S_p$. Both certified procedures use the weighted BY fallback at $q^\circ$, and direct weighted BY uses the full testing score at $q$. All correlations are checked in each certificate. We also evaluate diagnostic weighted BH on the full testing scores. The split comparator uses 150 rather than 299 residual contrasts, so the comparison combines the effects of subject allocation and reference-score construction. The six paired mean-power differences between reuse and direct weighted BY form one HB family, and the six FDP means of certified reuse form another.

\begin{table}[!htbp]
\centering
\caption{Monte Carlo summaries under a common subject budget. Acc. denotes certificate acceptance frequency. The split comparator uses 300 testing subjects and 150 paired certification contrasts.}
\label{tab:fixed-budget-comparison}
\begingroup
\footnotesize
\setlength{\tabcolsep}{2.5pt}
\renewcommand{\arraystretch}{1.12}
\begin{tabular*}{\linewidth}{@{\extracolsep{\fill}}lr rrr rr rrr@{}}
\toprule
& & \multicolumn{3}{c}{Certified reuse} & \multicolumn{2}{c}{Direct wBY} & \multicolumn{3}{c}{Certified split} \\
\cmidrule(lr){3-5}\cmidrule(lr){6-7}\cmidrule(lr){8-10}
Regime & $s_J(\Gamma)$ & Acc. & FDP & Power & FDP & Power & Acc. & FDP & Power \\
\midrule
Mixed sign & -0.301 & 0.000 & 0.017 & 0.454 & 0.018 & 0.468 & 0.000 & 0.015 & 0.183 \\
Independence & 0.000 & 0.000 & 0.017 & 0.438 & 0.018 & 0.453 & 0.000 & 0.016 & 0.158 \\
Weak positive & 0.070 & 0.000 & 0.016 & 0.449 & 0.017 & 0.464 & 0.000 & 0.015 & 0.182 \\
Equicorr. $0.30$ & 0.300 & 0.641 & 0.046 & 0.595 & 0.018 & 0.467 & 0.000 & 0.015 & 0.177 \\
Heterogeneous positive & 0.413 & 1.000 & 0.058 & 0.677 & 0.017 & 0.473 & 0.080 & 0.019 & 0.201 \\
Equicorr. $0.50$ & 0.500 & 1.000 & 0.059 & 0.673 & 0.017 & 0.471 & 0.083 & 0.018 & 0.201 \\
\bottomrule
\end{tabular*}

\endgroup
\end{table}

The diagnostic weighted BH had mean FDP below $0.1$ in each of the six regimes. This fixed-budget study shows power recovery in the certified positive dependence cases, but it does not itself show a nominal-FDR violation of weighted BH. The analytical counterexample in \cref{ex:weighted-bh-inflation} addresses the latter question.

\FloatBarrier

\subsection{Sensitivity to changes in an HBN-derived Gaussian covariance}
\label{app:hbn-covariance-stress}

We use the frozen ten-dimensional correlation matrix $\Gamma_{\mathrm{HBN}}$ from \cref{app:hbn-bridge}. Let $D$ be a diagonal matrix that changes the sign of coordinate 9, a true null, and leaves the remaining coordinates unchanged. For $t\in[0,1]$, set
\begin{equation}\label{eq:hbn-perturbation}
\Gamma_t=(1-t)\Gamma_{\mathrm{HBN}}+tD\Gamma_{\mathrm{HBN}}D.
\end{equation}
Every $\Gamma_t$ is a valid correlation matrix because it is a convex combination of correlation matrices. The sequence progressively attenuates and then reverses the signs of correlations involving one null endpoint. It represents controlled covariance perturbations, not longitudinal or additional observational HBN data.

We simulated 30,000 replications at each $t\in\{0,.05,.10,.20,.50,.75,1\}$. The study sample contains 600 Gaussian subjects and the independent pilot contains 100 subjects. The first two subject means equal $2/\sqrt{600}$ and all others are zero. The pilot weights follow the construction in \cref{app:fixed-budget-study}. The full testing score and its independent certification scatter have distributions $N_{10}(\sqrt{600}\,\theta,\Gamma_t)$ and $\operatorname{Wishart}_{10}(599,\Gamma_t)$, respectively. We check all pairs, divide the scatter by 599, and compare Cert-wBY with direct weighted BY using the same weights and scores.

At $t=0$, the certificate accepted in $0.8653$ of replications, with mean power $0.4681$ versus $0.3216$ for direct weighted BY. At $t=0.10$, acceptance fell to $0.4445$ and the paired mean-power difference decreased to $0.0689$. By $t=0.20$, acceptance was $0.0685$ and the difference was $0.0004$. The seven paired differences form one HB family with simultaneous 95\% half-width $0.0194$.

\begin{figure}[!htbp]
\centering
\includegraphics[width=.92\linewidth]{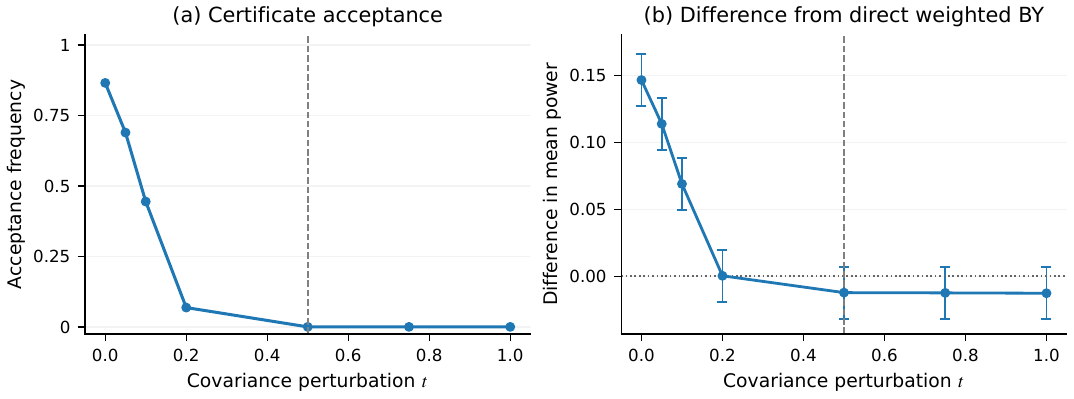}
\caption{Certificate acceptance (a) and paired mean-power difference relative to direct weighted BY (b) along \eqref{eq:hbn-perturbation}. Dashed vertical lines mark the sign boundary $t=0.5$. Error bars in (b) are simultaneous 95\% HB intervals.}
\label{fig:hbn-perturbation}
\end{figure}

\FloatBarrier

\subsection{HBN data construction and exploratory audit}
\label{app:hbn-provenance}

The exploratory analysis used functional-connectivity derivatives from the Healthy Brain Network \citep{alexander2017hbn}, distributed through the Reproducible Brain Charts HBN XCP-D resource \citep{reprobrainchart2026hbnxcpd}. The analyzed resources correspond to XCP-D 0.10.6 and fMRIPrep 25.0.0 derivatives \citep{reprobrainchart2026hbnxcpd,reprobrainchart2026hbnfmriprep}.
The XCP-D and fMRIPrep pipelines are described by \citet{mehta2024xcpd} and \citet{esteban2019fmriprep}, respectively. No restricted phenotypic variables entered the analysis. The XCP-D revision is identified in the cited repository record; the upstream fMRIPrep commit could not be verified.

\paragraph{Data access and use.} The HBN neuroimaging data are available without a data use agreement, unlike restricted phenotypic records \citep{hbn2026faq}. The original Healthy Brain Network study obtained institutional ethical approval and informed consent from participants or their legal guardians \citep{alexander2017hbn}. The present study uses existing imaging derivatives for noncommercial methodological research and does not involve participant recruitment or the collection of new participant data. HBN distributes its data under CC BY 4.0 or, for a subset of participants, a Creative Commons Attribution-NonCommercial-ShareAlike license. The participant-specific licensing metadata identify those subject to noncommercial restrictions \citep{hbn2026license}. No participant-level records are redistributed with this manuscript.

The archived preprocessing configuration specifies resting-state fMRI, Pearson correlation derivatives in fsLR space, and the Gordon cortical parcellation \citep{gordon2016generation}. The Gordon cortical atlas has a nominal resolution of 333 parcels. The Default network comprises 41 regions of interest(ROIs) in the feature bundles. The archived quality-control configuration required no motion-exclusion flag, median framewise displacement at most 0.2, 365--420 retained volumes, and coverage above 0.5 for every Default parcel. Its scan-selection rule prioritized the largest numeric run (a missing run was ranked as zero), then the largest retained-volume count, then descending lexicographic file name. The final scan-level manifest has not been independently reconciled with these settings, and the archived geometry and observed-data summaries have not been independently reconstructed from participant-level inputs. After retaining one scan per participant, the reported learning, certification, and testing splits contained 292, 1,365, and 292 participants, respectively.

\paragraph{Connectivity summaries.}
For participant \(s\) and Default-network ROI \(i\), let
\[
F_{si}
=
\frac1{40}
\sum_{k\ne i}
\operatorname{atanh}\{
\operatorname{clip}(r_{s,ik},-1+10^{-12},1-10^{-12})
\},
\]
where $r_{s,ik}$ is the participant-level Pearson correlation between ROIs $i$ and $k$, and the sum runs over the other 40 ROIs in the Default network.
Thus the hypotheses concern node-level mean Fisher-\(z\) connectivity
summaries, not individual seed--target edges.

Ten ROI summaries were selected by a prespecified ranking of their labels without using connectivity values. The coordinate order was fixed throughout the numerical study:
\[
\begin{split}
&\texttt{L\_Default\_20},
\texttt{R\_Default\_35},
\texttt{L\_Default\_17},
\texttt{L\_Default\_16},
\texttt{R\_Default\_31},\\
&\texttt{R\_Default\_36},
\texttt{R\_Default\_27},
\texttt{L\_Default\_9},
\texttt{R\_Default\_22},
\texttt{L\_Default\_12}.
\end{split}
\]

\paragraph{Adjustment and testing.}
Let $x_s$ consist of two acquisition-stratum indicators (with the standard stratum omitted), median framewise displacement, and the number of retained volumes, each centered at its learning-sample mean. For each selected feature, we fit a linear regression with an intercept using only $\mathcal D_W$. Let $\widehat b_i$ denote the four fitted slope coefficients and define
\begin{equation}\label{eq:hbn-adjustment}
\widetilde F_{si}=F_{si}-x_s^\top\widehat b_i.
\end{equation}
The scale $\widehat\sigma_i$ is the sample standard deviation of the adjusted feature in $\mathcal D_W$, with divisor $N_W-1$. The fitted slopes and scales are held fixed for the certification and testing samples.

For a reference value $r_0$, set $z_0=\operatorname{atanh}(r_0)$ and define the standardized subject-level score
\begin{equation}\label{eq:hbn-score}
V_{si}=\frac{\widetilde F_{si}-z_0}{\widehat\sigma_i}.
\end{equation}
Locally index the testing subjects by $s=1,\ldots,N_T$. The testing score and $p$-value are
\begin{equation}\label{eq:hbn-testing}
Z_i=N_T^{-1/2}\sum_{s=1}^{N_T}V_{si},
\qquad
P_i=1-\Phi(Z_i).
\end{equation}

The test in \eqref{eq:hbn-testing} therefore concerns the mean adjusted node-level Fisher-\(z\) summary relative to \(z_0\) and should not be interpreted as a test that the mean raw correlation exceeds \(r_0\).

Because the marginal scales are estimated, finite-sample
conditional standard-normal calibration of \(Z_i\) has not been established. The resulting \(p\)-values are therefore used only in the exploratory analysis.

\paragraph{Certification.}
Within each acquisition stratum, certification participants were permuted using a prespecified pseudorandom rule and grouped into consecutive disjoint pairs. If the stratum size was odd, the final participant was omitted. Write $V_s=(V_{s1},\ldots,V_{sm})^\top$ using \eqref{eq:hbn-score}.
For pair \((a_\ell,b_\ell)\),
\[
Y^{(\ell)}=\frac{V_{b_\ell}-V_{a_\ell}}{\sqrt2}.
\]

The cross-moment matrix is computed by \eqref{eq:certification-matrix}. The reference value $z_0$ cancels from each paired difference, so the cross-moment calculation is unchanged by this centering.

The three strata contributed 119, 82, and 481 disjoint pairs, giving
\(N_C=682\). Thus 1,364 of the 1,365 certification participants entered the paired cross-moment calculation. The remaining participant belonged to the standard stratum and was not used for certification.

This matrix averages paired cross-products and is not renormalized. Under \cref{ass:gaussian-certification}, its conditional expectation is $\Gamma$ and its diagonal elements have the distribution $\chi^2_{N_C}/N_C$. A realization of the matrix need not have diagonal entries equal to one. In observed data, the adjustment estimated in the learning sample need not yield centered paired differences or unit marginal variances. In particular, if the paired subjects have different conditional means, the cross-moment includes an additional product of mean differences. Thus the Gaussian calibration of the estimator requires assumptions beyond the observed preprocessing.

\paragraph{Coverage and interpretation.}
The primary eligibility criterion required coverage greater than one half in every Default-network parcel. Complete coverage was not required. Among the retained participants, 4 in the learning sample, 29 in the certification sample, and 7 in the testing sample had incomplete coverage in at least one parcel. All retained feature vectors contained finite summaries for the 820 within-network edges and 41 nodes. A sensitivity analysis restricted to complete coverage was not performed.

Conditioning on \(\mathcal D_W\) fixes the fitted adjustment and scale
estimates but does not establish conditional Gaussianity or equality of
the certification and testing population correlation matrices.
Site-specific covariance may differ, and family identifiers were
unavailable, so separation of related participants across splits could
not be audited. The observed-data analysis therefore does not provide
finite-sample validation of \cref{thm:sign-certificate}.

At $r_0=0.1$, the minimum checked cross-moment was $0.2416$, exceeding the threshold $0.1779$. The certificate therefore selected the weighted BH branch. For weighting, index the learning subjects by $s=1,\ldots,N_W$ and set $Z_i^W=N_W^{-1/2}\sum_{s=1}^{N_W}V_{si}$. We use \eqref{eq:hbn-weights} with $\eta=1$. The learned weights were uniform, and all compared procedures rejected all ten hypotheses. Consequently, this analysis does not distinguish their comparative performance.

\FloatBarrier

\subsection{Computational details}
\label{app:computational-details}

For $L$ prespecified Monte Carlo means, each based on $B$ independent replications of a variable $U\in[a,b]$, let $\bar U$ denote its sample mean and set
\begin{equation}\label{eq:hb-halfwidth}
h=(b-a)\sqrt{\frac{\log(2L/0.05)}{2B}}.
\end{equation}
Hoeffding's inequality gives
\[
\Pr(|\bar U-\mathbb E U|>h)
\le 2\exp\{-2Bh^2/(b-a)^2\}=0.05/L.
\]
The union bound gives simultaneous coverage of at least $0.95$.

We used \([a,b]=[0,1]\) for FDP and power and \([-1,1]\) for paired
power differences. Intervals were truncated to the corresponding
parameter range. Certificate acceptance in the HBN Gaussian simulation used the two-sided 95\% Clopper--Pearson interval.

\begin{table}[H]
\centering
\captionsetup{font=small,skip=8pt}
\caption{Computational details for the numerical study.}
\label{tab:computational-details}
\small
\renewcommand{\arraystretch}{1.15}
\begin{tabular}{@{}p{0.32\linewidth}
p{\dimexpr0.68\linewidth-2\tabcolsep\relax}@{}}
\toprule
Item & Specification \\
\midrule
Routing transition & $200{,}000$ per original grid point and $50{,}000$ per refined grid point \\
Certificate calibration & $200{,}000$ per configuration \\
Moderate-dimensional stress & $200{,}000$ per configuration \\
HBN Gaussian simulation & $100{,}000$ per $(\nu,\eta)$ \\
Fixed budget and covariance sensitivity & $20{,}000$ and $30{,}000$ per configuration, respectively \\
Randomness & Independent Monte Carlo replications within each configuration. Competing procedures share the draws specified in their designs. \\
\bottomrule
\end{tabular}
\end{table}
\FloatBarrier

\end{document}